\documentclass[aps,pra,twocolumn,showpacs,superscriptaddress,10pt]{revtex4-1}

\usepackage{amsmath,bm,amssymb}
\usepackage{graphicx,xcolor,soul}
\usepackage{tabularx,ragged2e,booktabs}
\usepackage[colorlinks=true,linkcolor=blue,urlcolor=cyan,citecolor=red,anchorcolor=blue]{hyperref}

\begin{document}

\title{Inversion of Strong Field Photoelectron Spectra for Molecular Orbital Imaging}

\author{R. Puthumpally-Joseph}
\affiliation{Institut des Sciences Mol\'eculaires d'Orsay (ISMO), CNRS, Univ. Paris-Sud, Universit\'e Paris-Saclay, F-91405 Orsay (France)}

\author{J. Viau-Trudel}
\affiliation{Institut des Sciences Mol\'eculaires d'Orsay (ISMO), CNRS, Univ. Paris-Sud, Universit\'e Paris-Saclay, F-91405 Orsay (France)}
\affiliation{D\'epartement de Chimie, Universit\'e Laval, Qu\'ebec, Canada G1K 7P4}

\author{M. Peters}
\affiliation{Universit\'e de Moncton, Edmundston, NB, Canada E3V 2S8}

\author{T. T. Nguyen-Dang}
\affiliation{D\'epartement de Chimie, Universit\'e Laval, Qu\'ebec, Canada G1K 7P4}

\author{O. Atabek}
\affiliation{Institut des Sciences Mol\'eculaires d'Orsay (ISMO), CNRS, Univ. Paris-Sud, Universit\'e Paris-Saclay, F-91405 Orsay (France)}

\author{E. Charron}
\affiliation{Institut des Sciences Mol\'eculaires d'Orsay (ISMO), CNRS, Univ. Paris-Sud, Universit\'e Paris-Saclay, F-91405 Orsay (France)}

\begin{abstract}
Imaging structures at the molecular level is a fast developing interdisciplinary research field that spans across the boundaries of physics and chemistry. High spatial resolution images of molecules can be obtained with photons or ultrafast electrons. In addition, images of valence molecular orbitals can be extracted via tomographic techniques based on the coherent XUV radiation emitted by a molecular gas exposed to an intense ultra-short infrared laser pulse. In this paper, we demonstrate that similar information can be obtained by inverting energy resolved photoelectron spectra using a simplified analytical model.
\end{abstract}
\maketitle

%%%%%%%%%%%%%%%%%%%%%%%
\section{Introduction}%
\label{Sec:Intro}     %
%%%%%%%%%%%%%%%%%%%%%%%

Assessing geometric and electronic structures of molecules via different scattering techniques using x-rays, ultrafast electrons and high harmonic generations (HHG) is a hot topic of current research in molecular physics since they provide a gateway to image chemical reactions in real time\;\cite{PhysRevLett.114.255501, PhysRevLett.109.133202}. Conventional scattering techniques based on photons and electrons are able to achieve spatial resolutions needed for imaging static molecular geometry but they lack resolution in time to give a dynamic picture. Techniques based on  strong field ionization and ultrafast lasers are promising as they can be used to provide both sub-\AA ngstrom spatial and sub-femtosecond temporal resolutions\;\cite{JPhysB.49.062001, RevModPhys.72.545} for dynamic imaging purposes. It is one of these techniques associated with strong-field ionization of molecules that is of interest in the present paper. For a thorough account of the latest trends in ultrafast molecular imaging methods, we refer the reader to Ref.\;\cite{DiMauro_et_al_JPhysB}.

The currently accepted vision of strong field ionization is the celebrated three-step model\;\cite{PhysRevLett.71.1994, PhysRevA.49.2117}. When an atom or a molecule is excited with an intense infrared (IR) laser pulse, a quasi static potential barrier is formed in the combined potential curve of the system and the field through which a bound electron can tunnel out\;\cite{jphysb47.204001}. It creates a laser driven electron wave packet in the ionization continuum, which is driven back and forth to the parent core by the applied field. On its return to this ionic core, the electron wave packet is scattered, resulting either in elastic scattering or in inelastic collision processes like high harmonic generation (HHG) or non-sequential double ionization (NSDI)\;\cite{Zuo1996313, JPhysB21.l31, PhysRevA.48.R894} for instance.

Compared to the relatively inefficient inelastic processes, the elastic scattering of the ionized wave packet is the predominant outcome of the recollision. It is known as Laser Induced Electron Diffraction (LIED). Following the first theoretical discussions in 1996\;\cite{Zuo1996313}, experimental realization of LIED on simple molecular systems was first reported in 2008\;\cite{Meckel13062008}. Since then, LIED has been considered as a tool to study strong field dynamics of isolated molecules.

It is well-known from optical physics that a diffraction pattern can be seen as the image of an object in the reciprocal space, from which light, or an incident matter wave, has been scattered. By designing an inverse algorithm, one can reconstruct the image of that object in real space. Laser induced electron diffraction can be seen in the same perspective\;\cite{Zuo1996313, Meckel13062008} but unlike traditional scattering processes, in LIED the scattering beam of electrons is extracted from the molecular system itself, acting as its own electron gun. After ionization and after an eventual recollision event, the outgoing electron wave carries information about the scattering centers. The photoelectron spectrum can thus be considered as an image of the system in the reciprocal space. Given the similarity of LIED with traditional diffraction techniques, it seems potentially possible to get information about the molecule from its LIED photoelectron spectra.

It was demonstrated both theoretically and experimentally that LIED can be used for extracting structural information about the equilibrium geometry of molecules with great accuracy\;\cite{Pullen2015, nat.483.194, PhysRevA.85.053417}. These recent experiments motivate developing LIED-based techniques for imaging molecular dynamics. In particular, experimental developments reported in Ref.\;\cite{Pullen2015} demonstrate the simultaneous measurement of both C-H and C-C bond lengths of aligned C$_2$H$_2$ using LIED spectra obtained with mid-IR laser fields. That the LIED spectral data can be inverted to retrieve precise information on the molecular geometry is not surprising, although it undoubtedly represents a huge advance in molecular physics, given that  this measure can in principle be made on a very short time scale, allowing molecular geometry changes during a reactive collision, for example, to be followed in a time-resolved manner.

Recently, one of the key research topic in strong field physics appeared to be the exploitation of the recollision process to retrieve not only structural or geometrical images but also to infer information on the electronic charge distribution of a molecule and even details of its field-free quantum eigenstates. Those pieces of information are of great interest, especially for the understanding and the imaging of reaction dynamics, where the changes in the electronic charge distribution play a major role. Achieving required spatial and temporal resolution could provide a tool for probing the transition states of a chemical reaction for example, by observing time-resolved deformation of the orbitals as transition states are crossed. Such a tool would also be of a tremendous value to image the rapid dynamics which takes place close to conical intersections\;\cite{jphyschemA.102.4031, annurev-physchem-032210-103522}.

Currently, HHG is the only strong field process that has been explored as a tool for imaging molecular orbitals using tomographic techniques\;\cite{nat.7.822}, as originally demonstrated in\;\cite{nat.432.867}. This  HHG-based orbital imaging approach involves a rather  elaborated inversion procedure, requiring the HHG spectra to be recorded at various laser-molecule alignment angles  and their treatment, \textit{i.e.} the inversion procedure per-se, rests on a number of assumptions that are still a matter of debate.

In this paper, we propose an alternative route that can be used to extract both structural and orbital information of a molecule directly from its LIED spectra. Previously, we demonstrated how LIED signals, for a symmetric molecule such as CO$_2$, reflect the conservation of the nodal structure, \textit{i.e.} the symmetry character, of the initial molecular orbital (MO) from which the ionized electron has been extracted. Here, we will show that more detailed information on this initial orbital can be retrieved from this signal, culminating with an explicit, complete MO reconstruction procedure.

The outline of the paper is as follows. In Sec.\;\ref{Sec:Model} we briefly recall the single-active electron (SAE) model of the CO$_2$ molecule as defined in the previous work and used in the present study, together with the numerical procedure for electron wave packet calculations within this model. Then, in Sec.\;\ref{Sec:Num_Res}, through results of numerical simulations, we illustrate the specific features of the photoelectron LIED spectrum associated with a \textit{molecular} orbital compared to the one of a typical \textit{atomic} orbital. In Sec.\;\ref{Sec:SFA_mod}, we derive an analytical expression of the LIED photoelectron momentum distribution, starting from formally exact integral expressions of the time-evolution operator describing the SAE dynamics. The final analytical model makes use of the  strong field approximation (SFA) and the inversion procedure used for the MO reconstruction  assumes a simple LCAO expression as a guess for the initial MO. Finally in Sec.\ref{Sec:Reconstruction}, we demonstrate this procedure in the case of the highest occupied molecular orbital (HOMO) of the carbon dioxide molecule. We present some examples of reconstruction and we specify the accuracy and limits of our approach. The last section gives some concluding remarks and perspectives for future work. Atomic units are used throughout the paper unless stated otherwise.

%%%%%%%%%%%%%%%%%%%%%%%%%%%%
\section{Theoretical model}%
\label{Sec:Model}          %
%%%%%%%%%%%%%%%%%%%%%%%%%%%%

To demonstrate how molecular orbitals can be imaged using LIED, we consider the specific case of the symmetric, linear, carbon dioxide molecule, CO$_2$, one of the most studied system in strong field physics\;\cite{PhysRevLett.98.243001, PhysRevA.83.051403, 0953-4075-27-7-002, Elshakre201337, doi:10.1021/ja0344819}. It is sufficiently complex to represent an interesting test case and it is relatively simple for calculations. It enables one to demonstrate the key features of electron dynamics in the presence of intense NIR fields\;\cite{PhysRevA.85.053417}.

The electronic dynamics induced by the field is described by the time-dependent Schr\"o\-dinger equation (TDSE)
\begin{equation}
\hat{\mathcal{H}}(t)\,|\psi(t)\rangle = i\,\partial_{t}\,|\psi(t)\rangle\, ,
\label{Eq:TDSE}
\end{equation}
where $|\psi(t)\rangle$ denotes the time-dependent electronic state of the model system constituted of the most weakly bound electron of the molecule and
\begin{equation}
\hat{\mathcal{H}}(t) =  - \bm{\nabla}^{2}/2 + V(\bm{r}) - \bm{\mu}\cdot\bm{E}(t)
\label{Eq:Hamiltonian_1}
\end{equation}
is its Hamiltonian in the length gauge. Here $V(\bm{r})$ is an effective field-free binding potential and $-\bm{\mu}\cdot\bm{E}(t)$ is the interaction of the active electron with the laser field. The linearly polarized electric field along $\hat{\bm{e}}_x$ is defined as
\begin{equation}
\bm{E}(t) =  - \partial_t\,\bm{A}(t)\,,
\label{Eq:E_of_t}
\end{equation}
where $\bm{A}(t)$ is the vector potential given by
\begin{equation}
\bm{A}(t) = \frac{E_0}{\omega_L}\,f(t)\,\cos(\omega_L t + \phi)\,\hat{\bm{e}}_x\,.
\label{Eq:Vect_pot}
\end{equation}
$\omega_L$ is the IR career frequency and $E_0$, the electric field amplitude. $\phi$ is the Carrier-Envelop Phase (CEP) and
\begin{equation}
f(t) = \sin^2\left(\frac{\pi t}{2\tau}\right)
\label{Eq:Envelop}
\end{equation}
denotes the temporal envelop of the pulse of Full Width at Half Maximum (FWHM) $\tau$.

The effective multi-well potential $V(\bm{r})$ is as given in\;\cite{PhysRevA.85.053417}. It is a soft Coulomb potential describing the attraction exerted on the single electron of the model system by  screened nuclear charges with a screening factor that, for each nucleus,  varies slowly with the distance separating the electron from the nuclear charge.   We assume that the CO$_2$ molecule is pre-aligned along the $y$ direction. The intense IR laser pulse given by Eq.\,(\ref{Eq:E_of_t}) is therefore applied normal to the molecular axis. Thus the ionization and associated dynamics are assumed to take place in the plane defined by the orthogonal system of coordinates consisting of the molecular $y$-axis and of the polarization $x$-axis of the applied time-dependent electric field.

\begin{figure}[ht]
\includegraphics[width=8.5cm]{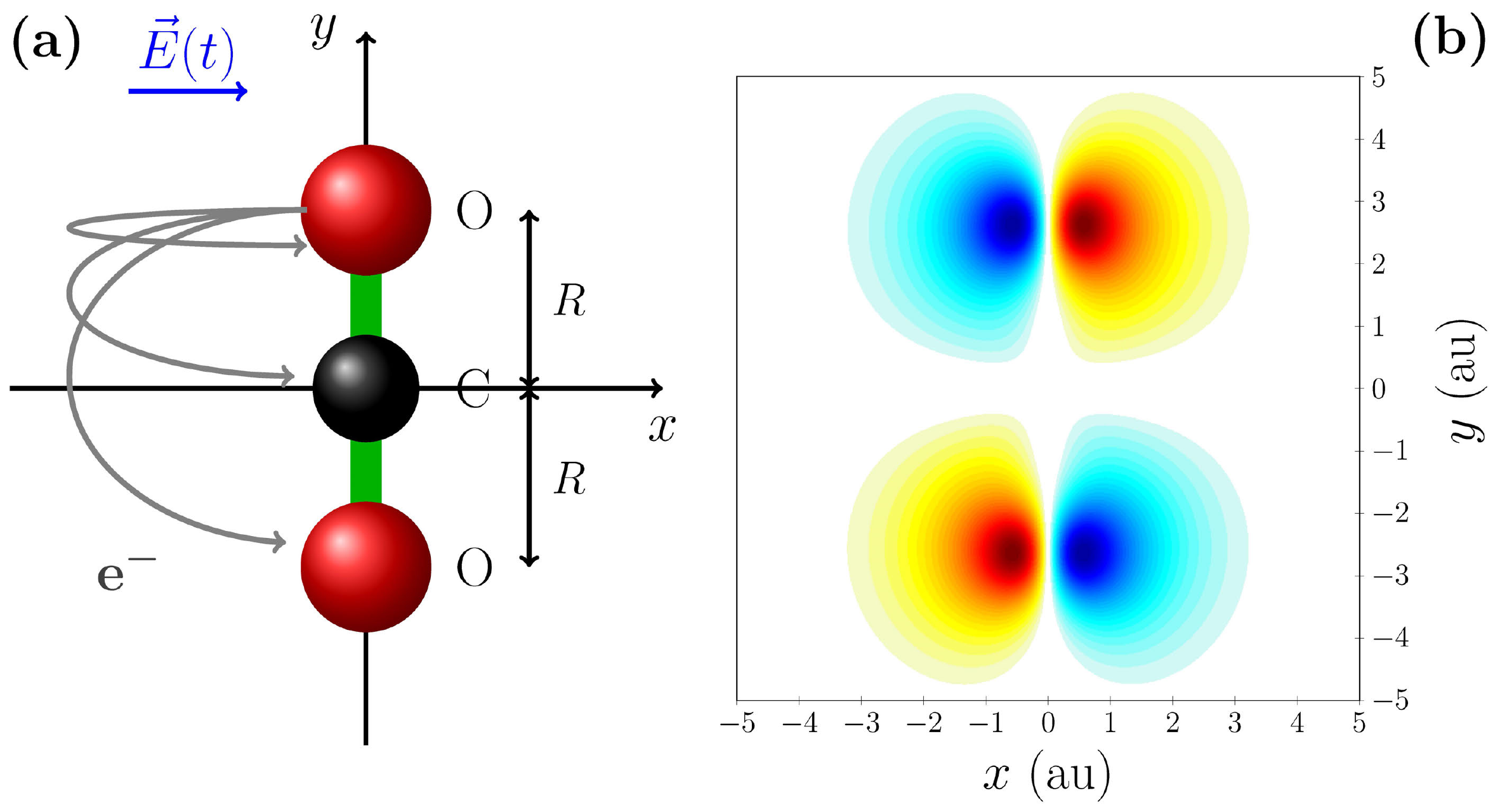}
\caption{(Color online).  (a) Model system with typical recollision trajectories. (b) HOMO wave function of a symmetric CO$_2$ molecule with the CO internuclear distance $R=R_e\simeq 1.4\,$\AA\,$\simeq 2.6\,$au.}
\label{fig:system}
\end{figure}

Fig.\;\ref{fig:system}\,(a) depicts the geometry of the system within these assumptions and shows in a schematic way three typical ionization and recollision trajectories. The most probable recollision processes take place following a short trajectory\;\cite{RevModPhys.81.163}, in about half an optical cycle, and therefore on a time scale of the order of 1 to 3 femtoseconds for wavelengths between 800\,nm and 2\,$\mu$m. The electronic dynamics that takes places on this typical time scale can be separated from the nuclear dynamics whose time scale is of the order of 15\,fs for the asymmetric stretch,  25\,fs for the symmetric stretch and of 60\,fs for the bending modes of CO$_2$. We therefore consider, in a first approximation, that the nuclear motion is frozen with a fixed CO bond length $R$.

The TDSE\;(\ref{Eq:TDSE}) describing the electronic dynamics is solved with the split-operator method\;\cite{QUA:QUA51}. The initial state is calculated using the imaginary time propagation technique\;\cite{Lehtovaara2007148} and the ionization and recollision events are simulated by propagating the calculated initial state during the pulse. During the interaction with the field, the asymptotic part of the wave packet is extracted and projected onto Volkov states in order to describe analytically the long range electronic dynamics\;\cite{PhysRevA.52.1450}. At the end of the pulse, corresponding to the time $t=t_f=2\tau$, the asymptotic part of the wave packet is collected to obtain the energy-resolved transition amplitudes and hence the photoelectron spectrum. The entire numerical procedure is detailed in\;\cite{PhysRevA.85.053417}. The calculated photoelectron spectrum is the laser-induced electron diffraction spectrum or LIED spectrum $\mathcal{I}(k_x,k_y)$, which gives the two dimensional momentum distribution of the elastically scattered electron wave packets.

%%%%%%%%%%%%%%%%%%%%%%%
\section{LIED Spectra}%
\label{Sec:Num_Res}   %
%%%%%%%%%%%%%%%%%%%%%%%

We discuss here the salient features of typical LIED spectra in preparation for the derivation of the inversion procedure of the next section. These spectra are calculated for the HOMO orbital of CO$_2$, seen in Fig.\;\ref{fig:system}(b), as the initial state. We also consider the spectra associated with the ionization out of a $2p_x$ atomic orbital centered on the carbon atom. This will be referred as the `atomic' case.

%%%%%%%%%%%%%%%%%%%%%%%%%%%%%%%%%%%%%%%%%
\subsection{Influence of the wavelength}%
%%%%%%%%%%%%%%%%%%%%%%%%%%%%%%%%%%%%%%%%%

\begin{figure}[ht]
\includegraphics[width=7.5cm]{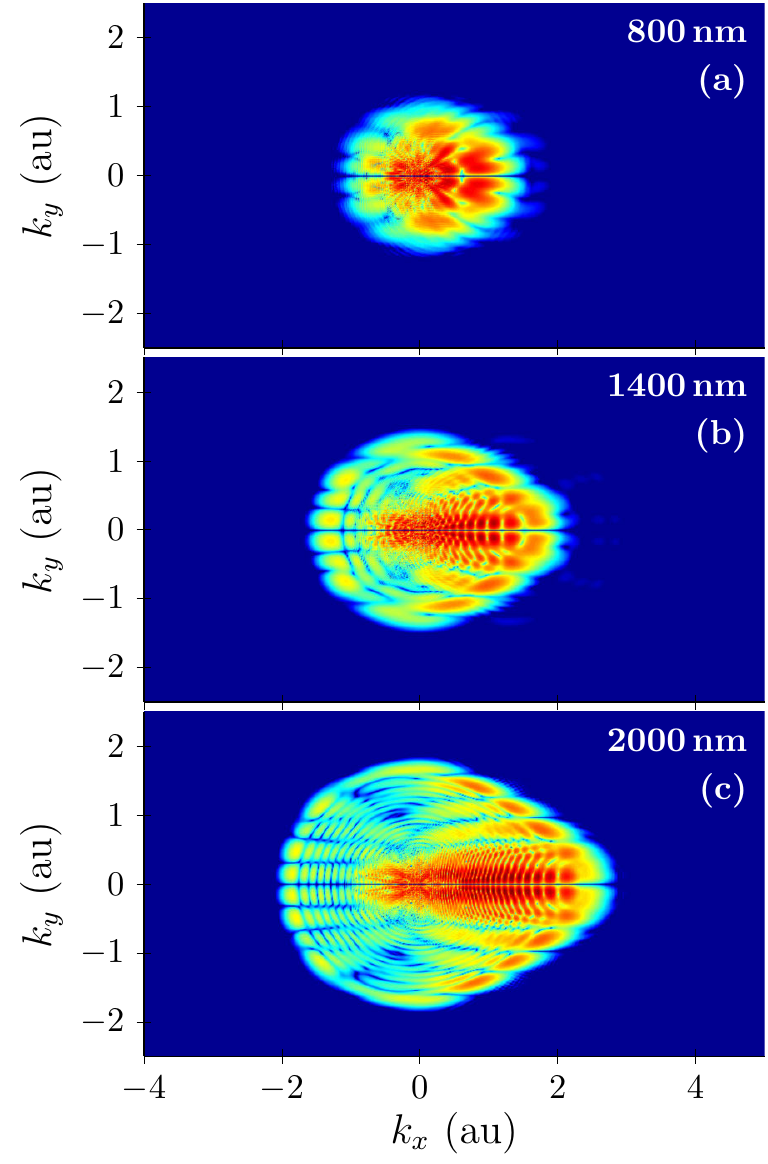}
\caption{(Color online). 2D photoelectron spectra $\mathcal{I}(k_x,k_y)$ (log scale) obtained from the HOMO of CO$_2$ for $R=5\,$\AA\/ when exposed to a single optical cycle pulse of intensity $I=10^{14}\,$W/cm$^2$ and zero CEP. The wavelengths used are (a) $\lambda=800\,$nm, (b) $\lambda=1.4\,\mu$m and (c) $\lambda=2.0\,\mu$m.}
\label{fig:spectra_lambda}
\end{figure}

LIED photoelectron spectra provide a picture of the momentum ($\bm{k}$) distribution of the ionized electron. A typical photoelectron spectrum $\mathcal{I}(k_x,k_y)$ obtained from the solution of the TDSE for the HOMO orbital of CO$_2$ at an extended geometry $R=5\,$\AA\/ is given in log scale in Fig.\;\ref{fig:spectra_lambda} for three different wavelengths and a single optical cycle pulse $(2\tau=2\pi/\omega_L)$ with no CEP $(\phi=0)$. Panel (a) shows the spectrum at the wavelength 800\,nm, panel (b) at 1.4\,$\mu$m and panel (c) at 2.0\,$\mu$m for a laser intensity of $10^{14}\,$W/cm$^2$. The highest probabilities are in red and the lowest in blue.

The outermost contour of the circular shape of the spectrum is elongated along $k_x$, \textit{i.e.} in the direction of the polarization of the field. Two successive ionization events corresponding to the maximum and minimum of $E(t)$ in this ultra-short pulse create an oscillating continuum wave packet which is ultimately driven away from the molecule. The ionization events happen along the direction of the field, giving photoelectrons with momenta  distributed as shown in the figure. The circular shape corresponds to the maximum recollision energy $3.17\,U_p= (k_x^2+k_y^2)/2 $, where $U_p$ is the ponderomotive energy\;\cite{RevModPhys.81.163}. Since $U_p$ is proportional to $\lambda^2$, an increase of the wavelength directly increases the size of the 2D photoelectron spectrum, as we can see in Fig.\;\ref{fig:spectra_lambda}. Longer wavelengths thus help making out the interference patterns of the spectrum. In the following we will use the largest wavelength $\lambda=2.0\,\mu$m.

%%%%%%%%%%%%%%%%%%%%%%%%%%%%%%%%%%%
\subsection{Interference patterns}%
%%%%%%%%%%%%%%%%%%%%%%%%%%%%%%%%%%%

\begin{figure}[ht]
\includegraphics[width=7.5cm]{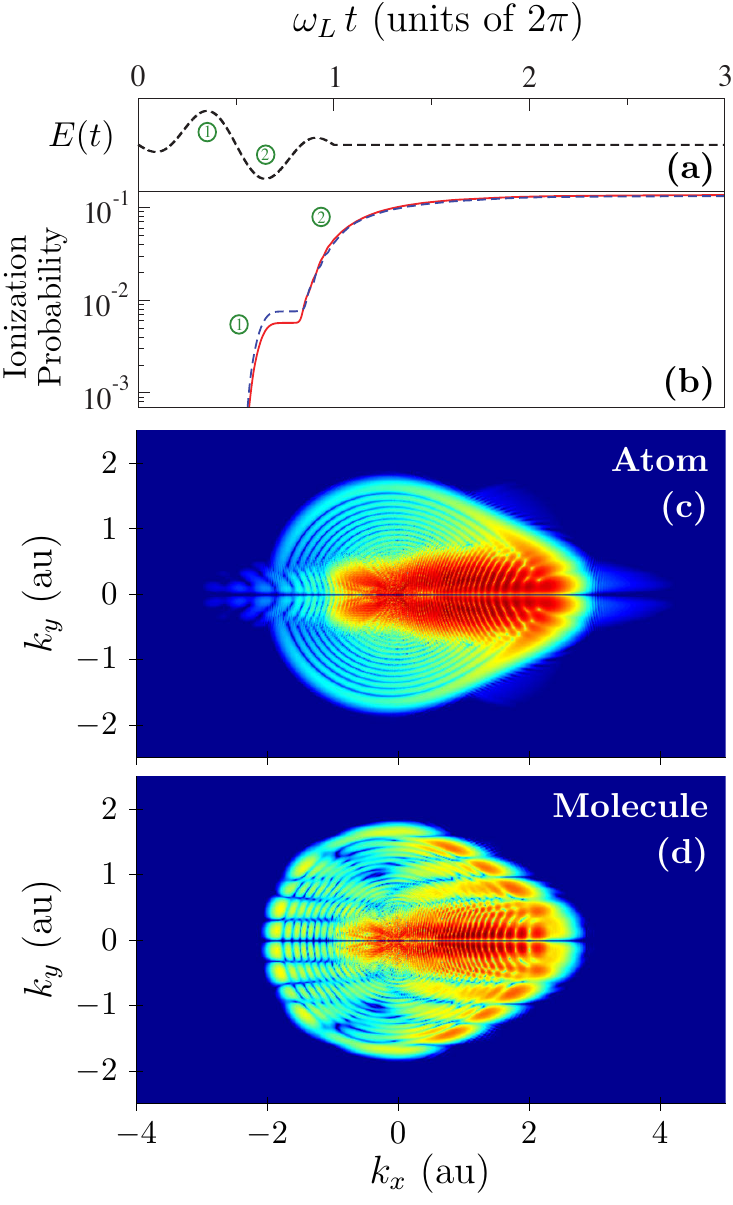}
\caption{(Color online). (a) Normalized electric field $E(t)$ as a function of time. (b) Ionization probability as a function of time for an atom (solid red line) and a molecule (dashed blue line) with the same ionization potential IP$=9.2\,$eV. (c) and (d) Associated 2D photoelectron spectra $\mathcal{I}(k_x,k_y)$ for the atom (c) and the molecule at $R=5\,$\AA\/ (d). A single optical cycle pulse of intensity $I=10^{14}\,$W/cm$^2$ and wavelength $\lambda=2.0\,\mu$m is used.}
\label{fig:mol_atom_1_opt}
\end{figure}

To analyze in detail the interference patterns which build up in the photoelectron spectra, we compare in panels (c) and (d) of Fig.\;\ref{fig:mol_atom_1_opt} the spectrum obtained from a $2p_x$ atomic orbital centered on the carbon atom with the spectrum obtained from the HOMO of CO$_2$, at a wavelength of 2.0\,$\mu$m. All other parameters are as in Fig.\;\ref{fig:spectra_lambda}. The panels (a) and (b) of the same figure show respectively the time variations of the electric field and of the total ionization probability for the atomic (red solid line) and for the molecular (dashed blue line) cases. For the atomic calculation, the parameters of the soft-core potential $V(\bm{r})$ in Eq.\;(\ref{Eq:Hamiltonian_1}) have been modified such that the atom has the same ionization potential compared to the HOMO of CO$_2$, \textit{i.e.} 9.2\;eV at $R=5\,$\AA. The electric field $E(t)$ presents two main symmetric maxima pointing in opposite directions. For both the atomic and the molecular cases, the ionization takes place in two successive bursts. The probability of ionization rises just after each maximum of the field, and the delay separating a maximum of the field and the associated ionization burst is simply related to the time necessary for the ionized wave function to reach the asymptotic region.

In the atomic photoelectron spectrum shown panel (c) very clear ring-like structures can be seen, which come from the interference between different rescattered electron wave packets. More precisely these structures are due to the interference between long and short trajectories followed by recolliding electrons\;\cite{Spanner:2004:12L02}. They have a circular shape because, for a given energy long and short trajectories accumulate a fixed phase shift which is independent of the electron emission angle.

Another interesting interference in the atomic LIED spectrum is due to the superposition of the pathways corresponding to direct ionization and to ionization preceded by recollision (\textit{i.e.} to rescattering). This holographic interference of the electron wave occurs only over a window of small $k_y$ values due to the limited spread of directly ionized electrons in the transverse direction\;\cite{Meckel2014, Chen:14, Huismans07012011}. It also appears mainly in the $k_x>0$ region of Fig.\;\ref{fig:mol_atom_1_opt}(c) (intense red colored region) due to the particular field $E(t)$ seen in panel (a) which drags the electron in the positive direction during the recollision. The associated interference patterns are relatively localized, \textit{i.e.} limited in extension, and are therefore difficult to measure in an experiment. In addition, they are seen both in atomic and molecular cases, as we can see in the comparison with panel (d) and they are therefore not the best candidates for an analysis of the molecular structure.

There is however a very clear and important difference between the atomic and molecular spectra which lies in the $k_y$ variation of the spectra. Indeed, out of the different interference patterns seen in the molecular spectrum, a multiple-slit like interference can be distinguished in the $k_y$ momentum distribution. This multiple-slit like interference pattern is due to the scattering of the electron by the multi-well ionic potential describing the interaction with the nuclei. The molecular information, including the relative position of the nuclei, is therefore mainly imprinted in the $k_y$ momentum distribution, along the direction of the molecular axis. To get a simpler spectrum that we can more easily analyze, we average the electron signal $\mathcal{I}(k_x,k_y)$ over the $k_x$ momentum, keeping only the $k_y$ variation. This yields the averaged 1D LIED spectrum
\begin{equation}
\mathcal{S}(k_y) = \int \mathcal{I}(k_x,k_y) \, dk_x\,.
\label{Eq:1D_spec_Eqn}
\end{equation}
It was already demonstrated that the bond length $R$ can directly be measured from the fringe width seen in this 1D spectrum\;\cite{PhysRevA.85.053417}.

\begin{figure}[ht]
\includegraphics[width=8.5cm]{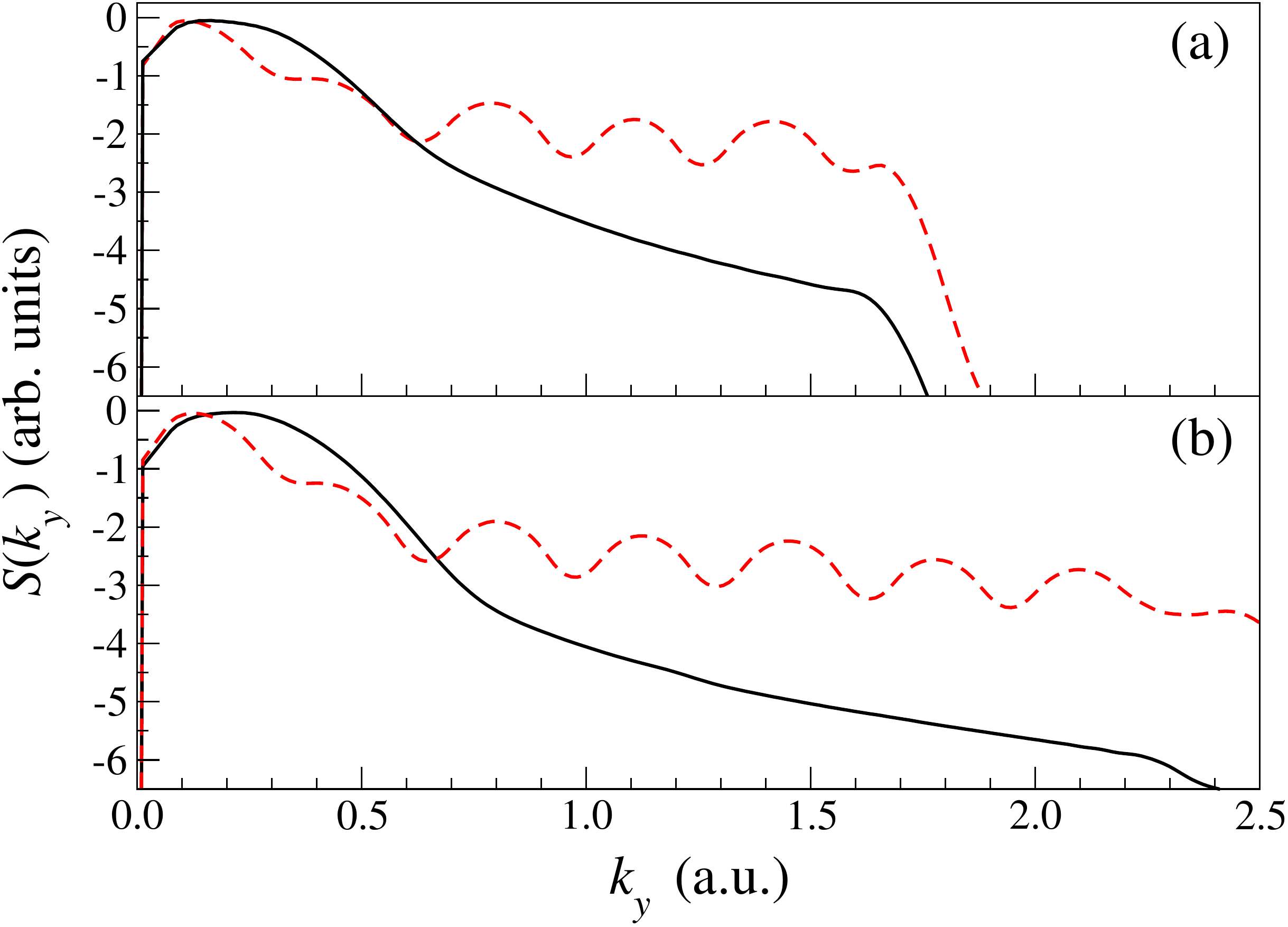}
\caption{(Color online). Averaged 1D LIED spectra $\mathcal{S}(k_y)$ (log scale) in the atomic case (solid black line) and in the molecular case (dashed red line). In panels (a) and (b) the parameters are as in Figs.\;\ref{fig:mol_atom_1_opt} and\;\ref{fig:mol_atom_4_opt}, respectively: (a) is for a total pulse duration of one optical cycle while (b) is for 3.5 optical cycles. All other parameters are identical.}
\label{fig:1DSpectrum}
\end{figure}

Two such log-scale spectra are shown in Fig.\;\ref{fig:1DSpectrum}(a) for the cases presented in 2D in Figs.\;\ref{fig:mol_atom_1_opt}(c) and\;\ref{fig:mol_atom_1_opt}(d). The averaged 1D atomic spectrum is shown as a solid black line and the molecular spectrum as a dashed red line. We clearly see strong differences in these 1D spectra which lie both in the oscillatory behavior of the molecular spectrum and in the slower decrease (with respect to $k_y$) of the mean signal of the molecular spectrum compared to the atomic spectrum.

\begin{figure}[ht]
\includegraphics[width=7.5cm]{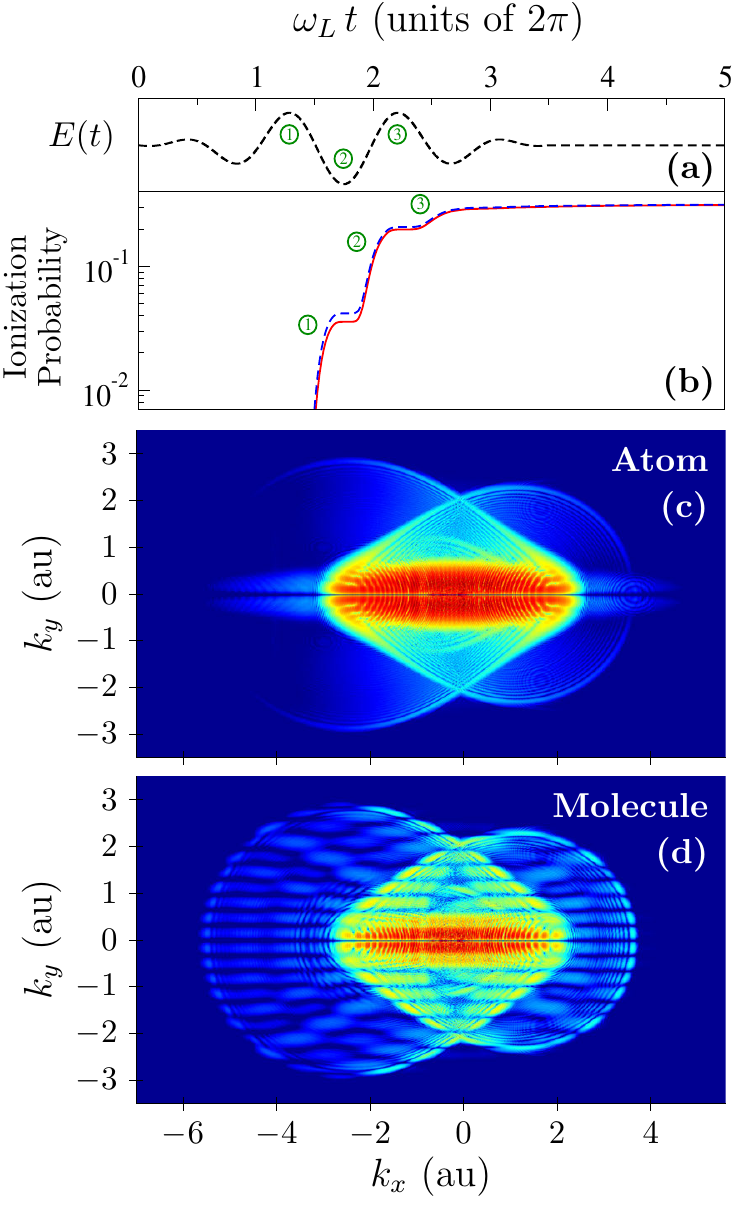}
\caption{(Color online). (a) Normalized electric field $E(t)$ as a function of time. (b) Ionization probability as a function of time for an atom (solid red line) and a molecule (dashed blue line) with the same ionization potential IP$=9.2\,$eV. (c) and (d) Associated 2D photoelectron spectra $\mathcal{I}(k_x,k_y)$ for the atom (c) and the molecule at $R=5\,$\AA\/ (d). A 3.5-optical cycle pulse characterized with an intensity of $I=10^{14}\,$W/cm$^2$ and the wavelength $\lambda=2.0\,\mu$m is used.}
\label{fig:mol_atom_4_opt}
\end{figure}

Until now, the LIED spectra were calculated for a single optical cycle only. Fig.\;\ref{fig:mol_atom_4_opt} shows similar atomic and molecular spectra, calculated with a 3.5-optical cycle laser pulse. Even though the pulse duration is much larger, there are only three main maxima of the electric field which contribute significantly to the ionization signal, as seen in panels (a) and (b) of Fig.\;\ref{fig:mol_atom_4_opt}. These maxima give rise to three bursts of ionization taking place in opposite directions. As a consequence the associated 2D momentum spectra are much more symmetric with respect to $k_x = 0$ than the spectra associated with a single-cycle pulse seen in Fig.\;\ref{fig:mol_atom_1_opt}.

The different kinds of interference patterns discussed above are still visible. In particular, the multiple-slit like interference seen in the $k_y$ variation of the 2D molecular spectrum is still present. The associated $k_x$-averaged 1D spectra seen in Fig.\;\ref{fig:1DSpectrum}(b) therefore show a similar behavior compared to the ultra-short single-cycle pulse. A comparison of Fig.\;\ref{fig:1DSpectrum}(a) and Fig.\;\ref{fig:1DSpectrum}(b) shows that the longer pulse yields a larger value of the cutoff energy. This is because the maximum value of $E(t)$ is larger for the longer pulse (see the panels (a) of Figs.\;\ref{fig:mol_atom_1_opt} and\;\ref{fig:mol_atom_4_opt}). The particular oscillatory behavior of $\mathcal{S}(k_y)$ in the molecular case of Fig.\;\ref{fig:1DSpectrum}(b) shows that it is possible to attempt an analysis of the molecular structure from LIED spectra using few-cycle laser pulses.

%%%%%%%%%%%%%%%%%%%%%%%%%%%%%%%%%%%%%%%%%%%%%%%%%%%%
\subsection{Influence of the internuclear distance}%
%%%%%%%%%%%%%%%%%%%%%%%%%%%%%%%%%%%%%%%%%%%%%%%%%%%%

\begin{figure}[ht]
\includegraphics[width=7.5cm]{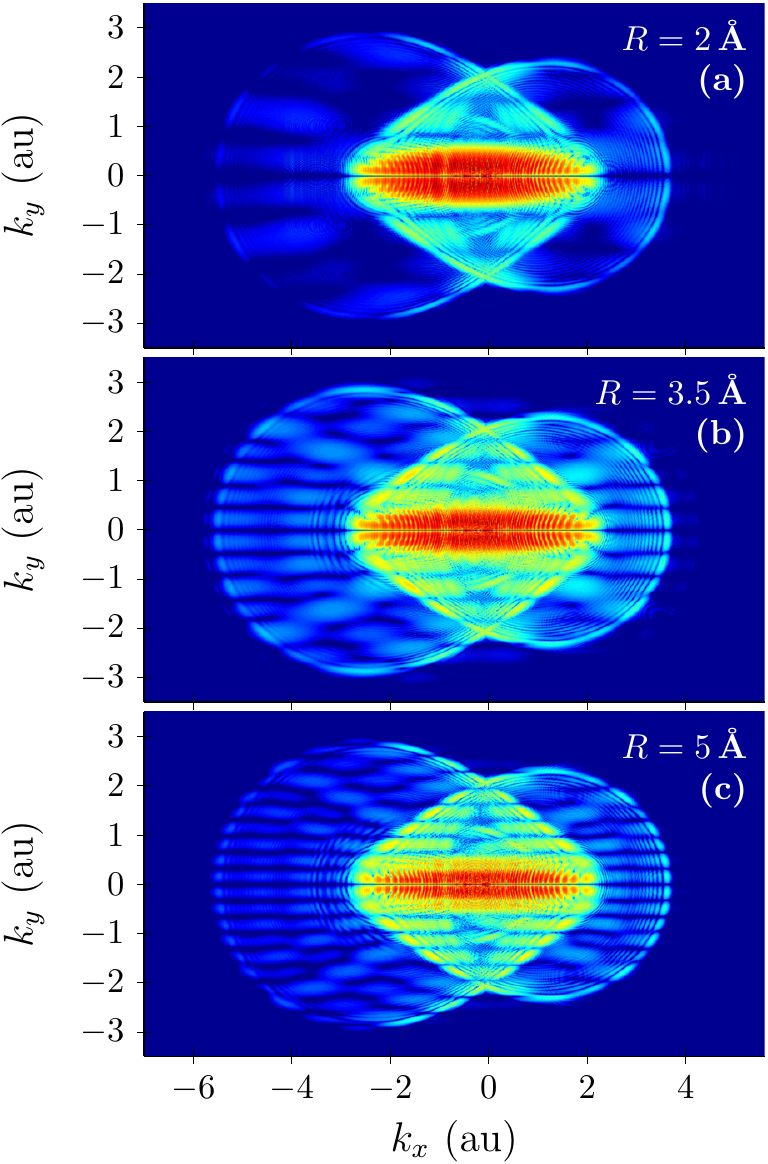}
\caption{(Color online). $R$-dependence of the 2D photoelectron spectra $\mathcal{I}(k_x,k_y)$ for the CO$_2$ molecule. The intensity is $I=10^{14}\,$W/cm$^2$ and the pulse duration is 3.5 optical cycles. The wavelength is $\lambda=2.0\,\mu$m. The internuclear distance is (a) $R=2\,$\AA, (b) $R=3.5\,$\AA\/ and (c) $R=5\,$\AA.}
\label{fig:spectra_R}
\end{figure}

In Fig.\;\ref{fig:spectra_R} we explore the $R$ dependence of the LIED spectra. The intensity is $10^{14}\,$W/cm$^2$ and the pulse duration is 3.5 optical cycles at the wavelength $2.0\,\mu$m. Panels (a), (b) and (c) are for $R=2.0$\,\AA, $R=3.5$\,\AA\/ and $R=5.0$\,\AA, respectively. We can conclude from this figure that the interference between long and short trajectories and the interference between direct ionization and ionization preceded by recollision (rescattering signal) are not seriously affected by a variation of the internuclear distance. On the other hand, the multiple-slit like interference patterns seen in the $k_y$ variation of the 2D molecular spectrum change appreciably when the internuclear distance varies.

\begin{figure}[ht]
\includegraphics[width=8.5cm]{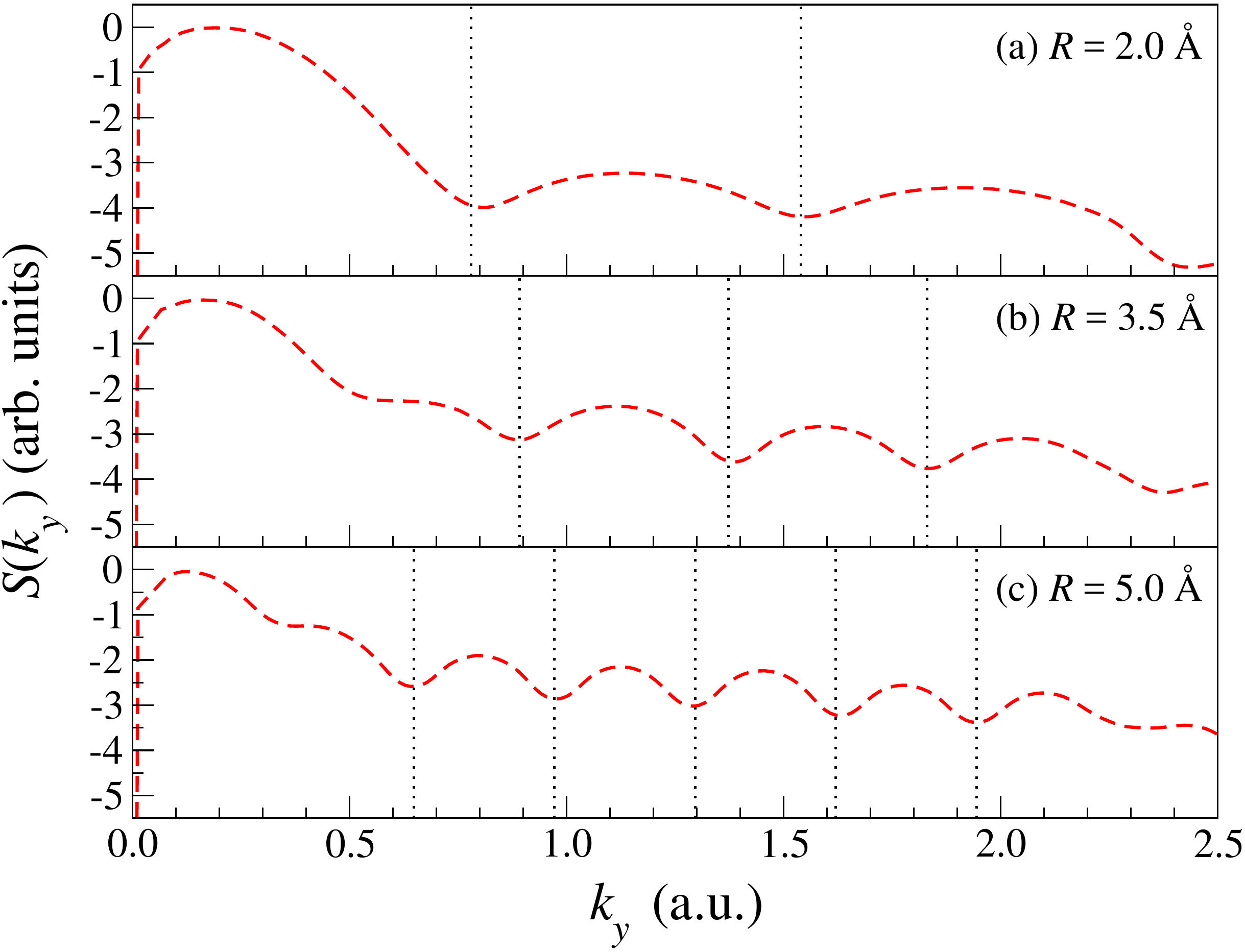}
\caption{(Color online). $R$-dependence of the averaged 1D LIED spectra $\mathcal{S}(k_y)$ (log scale). In panels (a), (b) and (c) the internuclear distance is $R=2.0$\,\AA, $R=3.5$\,\AA\/ and $R=5.0$\,\AA, respectively. The other parameters are as in Fig.\;\ref{fig:spectra_R}. The vertical dotted lines mark the regularly spaced local minima of the three different spectra.}
\label{fig:1DSpectrumR}
\end{figure}

This strong variation is confirmed by Fig.\;\ref{fig:1DSpectrumR} which shows the associated $k_x$-averaged one-dimensional LIED spectra. Panels (a), (b) and (c) are for $R=2.0$\,\AA, $R=3.5$\,\AA\/ and $R=5.0$\,\AA, respectively. We see here that the analysis of the spectrum is facilitated with large internuclear distances since the oscillation period of the 1D averaged spectrum decreases with $R$. Indeed, it was shown in\;\cite{PhysRevA.85.053417} that the fringe width $\Delta k$ varies as $\pi/R$. This result will be used in section\;\ref{Sec:Reconstruction} for the reconstruction of the initial molecular orbital.

For the laser parameters used in the present calculation, \textit{i.e.} $I=10^{14}\,$W/cm$^2$ and $\lambda=2.0\,\mu$m, the ponderomotive energy is $U_p=1.38$\,a.u, and, as seen in Fig.\;\ref{fig:spectra_R}, the electron spectrum extends over a range of momenta of a few atomic units only, with $k_y \leqslant 2.95$\,a.u. As we can already infer from Fig.\;\ref{fig:1DSpectrumR}(a) this range is not sufficient for an accurate analysis of the spectrum when $R < 3$\,a.u. In the following we will discuss this analysis for the cases $R=3.5\,$\AA\/ and $R=5.0\,$\AA. Analyzing the LIED spectra at smaller internuclear distances would require higher laser intensities or longer wavelengths.

The understanding the LIED spectra which was described in detail in this section can be used for the ultimate goal of this manuscript: the derivation of an inversion procedure. In the next section we describe the main ingredients of an analytical model that can lead to the image of the molecular orbital, in the present case the HOMO, by inverting the LIED spectrum. This model will then be used in the last section to analyze the spectra and to reconstruct the initial molecular orbital.

%%%%%%%%%%%%%%%%%%%%%%%%%%%%%%%%%%%%%%%%%%%%%%%%%%%
\section{The Inverse Problem: An Analytical Model}%
\label{Sec:SFA_mod}                               %
%%%%%%%%%%%%%%%%%%%%%%%%%%%%%%%%%%%%%%%%%%%%%%%%%%%

The 2D LIED spectrum $\mathcal{I}(k_x,k_y)$ calculated by solving the TDSE contains information about the molecule within the diffraction patterns, as described in Section\;\ref{Sec:Num_Res}. Since this spectrum originates from the HOMO orbital of CO$_2$, both structural and orbital information are necessarily imprinted in it. Here the goal is to reconstruct the initial orbital from which the photoelectrons are extracted. We are thus facing what could be called an \emph{inverse problem}, where we need a compact analytical form for the photoelectron spectra $\mathcal{S}(k_y)$, accurate enough to assess both orbital and geometrical information. This analytical form will contain some parameters describing the initial state. These parameters will be fitted such that the analytical form of $\mathcal{S}(k_y)$ reproduces its ``exact" counterpart obtained from the solution of the TDSE. Finally, the fitted parameters will be used to reconstruct the initial molecular orbital.

In general for the case discussed here, two main ingredients are necessary: (i)\;an approximate description of the ionization and associated dynamics that result in the photoelectron spectra and (ii)\;a simplified functional form for the initial state which will be used for the reconstruction. The first part is the most challenging feature of the inverse problem and is discussed in this section.

%%%%%%%%%%%%%%%%%%%%%%%%%%%%%%%%%%%%%%%%%
\subsection{Description of the Dynamics}%
%%%%%%%%%%%%%%%%%%%%%%%%%%%%%%%%%%%%%%%%%

%%%%%%%%%%%%%%%%%%%%%%%%%%%%%%%%%%%%%%%%%%%
\subsubsection{Exact Transition Amplitude}%
%%%%%%%%%%%%%%%%%%%%%%%%%%%%%%%%%%%%%%%%%%%

The field-induced dynamics can be modeled by depicting the different steps of a recollision event\;\cite{PhysRevLett.71.1994} separately. In agreement with this mechanism describing the ionization and recollision processes, we separate the transition amplitude $a(k_x,k_y)$ in two parts, corresponding to directly ionized electrons and to electrons ionized after a recolliding event.

If the exact solution $|\Psi(t_f)\rangle$ of the TDSE is known at the end of the pulse, at time $t_f$, the relevant transition amplitude for LIED can be written as
\begin{equation}
a(k_x,k_y) = \langle \Psi^{+}_{\bm{k}}|\Psi(t_f) \rangle\,,
\label{Eq:Analytic_tran_amp_1}
\end{equation}
where $|\Psi^{+}_{\bm{k}}\rangle$ is the outgoing wave elastically scattered in the direction of the electron wave vector $\bm{k}$ for a prescribed asymptotic kinetic energy $\varepsilon_k=k^2/2$. The formal solution of the TDSE may be written at time $t_f$ as
\begin{equation}
|\Psi(t_f)\rangle = \hat{U}(t_f \!\leftarrow\! 0)\,|\Psi(0)\rangle\,,
\end{equation}
where $|\Psi(0)\rangle$ is the initial state and $\hat{U}(t \!\leftarrow\! 0)$ is the evolution operator obeying the TDSE
\begin{equation}
i\,\partial_t\,\hat{U}(t \!\leftarrow\! 0) = \hat{\mathcal{H}}(t)\,\hat{U}(t \!\leftarrow\! 0)\,.
\label{Eq:evolution_op_TDSE}
\end{equation}
$\hat{\mathcal{H}}(t)$ given in Eq.\,(\ref{Eq:Hamiltonian_1}) contains both the binding and the driving potentials. Depending on the situation, one of them could be more influential than the other and could decide for the outcome of the dynamical process\;\cite{ATI_top_rev}.

The simplest realistic picture of strong field ionization including the essential ingredients of tunnel ionization followed by recollision, requires one to consider at least a complete optical cycle. For the derivation of the model, we therefore consider a single optical cycle of duration $t_f=2\pi/\omega_L$. For certain times $t'$ within this cycle, the field reaches values sufficient to trigger both tunnel ionization and the following dynamics of the wave packet, which can be represented using the following exact form of the Dyson equation\;\cite{Frasca2195, Reiss_SFA, PhysRevA.74.063404} :
\begin{eqnarray}
\hat{U}(t_f \!\leftarrow\! 0) & =    & \hat{U}_0(t_f \!\leftarrow\! 0) \label{Eq:Dyson_Eq_1} \\
                              & +i\, & \int_{0}^{t_f} \hat{U}(t_f\! \leftarrow\! t')\, \hat{\bm{\mu}}\cdot\bm{E}(t')\,\hat{U}_0(t' \!\leftarrow\! 0)\,dt'\,,\nonumber
\end{eqnarray}
where $\hat{U}_0(t \!\leftarrow\! 0)$ is the evolution operator associated with the field-free Hamiltonian
\begin{equation}
\hat{\mathcal{H}}_{0}=-\bm{\nabla}^2/2+V(\bm{r})\,.
\end{equation}

The Dyson equation (\ref{Eq:Dyson_Eq_1}) is exact in so far as it involves the exact evolution operator $\hat{U}(t_f\! \leftarrow\! t')$ between the time of ionization $t'$ and the final time $t_f$. During this time interval a recollision event may take place, whenever the electron wave packet propagating in the laser field comes close enough to the parent ionic core such that the Coulomb attraction starts to dominate over the driving dipole interaction. To express this idea, we then split the evolution operator $\hat{U}(t_f\! \leftarrow\! t')$, found in the integral on the right-hand-side (r.h.s.) of Eq.\,(\ref{Eq:Dyson_Eq_1}), as
\begin{eqnarray}
\hat{U}(t_f\! \leftarrow\! t') & =    & \hat{U}_v(t_f\! \leftarrow\! t') \label{Eq:Dyson_Eq_2} \\ \nonumber
                               & -i\, & \int_{t'}^{t_f} \hat{U}(t_f\! \leftarrow\! t'')\,V(\bm{r})\,\hat{U}_v(t''\! \leftarrow\! t')\,dt''\,,
\end{eqnarray}
where $\hat{U}_v$ is the evolution operator associated with the Volkov Hamiltonian\;\cite{Reiss_SFA, NDSenGupta}
\begin{equation}
\hat{\mathcal{H}}_v(t)=-\bm{\nabla}^2/2-\hat{\bm{\mu}}\cdot\bm{E}(t)\,.
\end{equation}

The Volkov evolution operator $\hat{U}_v(t_2 \leftarrow t_1)$ can be formally written as:
\begin{equation}
\hat{U}_v(t_2 \leftarrow t_1) = \int d\bm{k}\, \left|\Phi_{\bm{k}}^v(t_2)\right\rangle\left\langle\Phi_{\bm{k}}^v(t_1)\right|\,,
\label{Eq:volkov_projector_2D}
\end{equation}
where
\begin{equation}
\Phi_{\bm{k}}^v(\bm{r},t) = \frac{\mathrm{e}^{\,i\,\left[\bm{k}+ \,\bm{A}(t)\right]\,\cdot\,\bm{r}\,-\,i\,S(\bm{k},t)}}{2\pi}\,,
\label{Eq:volkov_states_2D}
\end{equation}
$S(\bm{k},t)$ being the classical action
\begin{equation}
S(\bm{k},t) = \frac{1}{2} \int_{0}^t \left[ \bm{k} + \,\bm{A}(\tau) \right]^2\,d\tau\,.
\end{equation}

Substituting Eq.(\ref{Eq:Dyson_Eq_2}) in Eq.(\ref{Eq:Dyson_Eq_1}) we get
\begin{equation}
\hat{U}(t_f\! \leftarrow\! 0) = \hat{U}_0(t_f) + \hat{U}_d(t_f) + \hat{U}_r(t_f) \,,
\label{Eq:Dyson_Eq_3}
\end{equation}
with the following definitions:
\begin{subequations}
\label{Eq:Evolution_Operators}
\begin{eqnarray}
\!\!\!\!\hat{U}_0(t_f) & = & \exp\big(-i\,\hat{\mathcal{H}}_0\,t_f\,\big)\,,        \label{Eq:field_free_Evolution}     \\[0.2cm]
\!\!\!\!\hat{U}_d(t_f) & = & i\int_{0}^{t_f}\!\!\!\!dt'\,\mathcal{D}(t_f,t')\,,  \label{Eq:direct_ionization_part_1} \\[0.2cm]
\!\!\!\!\hat{U}_r(t_f) & = & \int_{0}^{t_f}\!\!\!\!dt'\!\!\int_{t'}^{t_f}\!\!\!\!dt''\,\hat{U}(t_f\! \leftarrow\! t'')\,V(\bm{r})\,\mathcal{D}(t'',t')\,,
\end{eqnarray}
\end{subequations}
and
\begin{equation}
\mathcal{D}(t_2,t_1) = \hat{U}_v(t_2 \leftarrow t_1)\,\hat{\bm{\mu}}\cdot\bm{E}(t_1)\,\hat{U}_0(t_1 \leftarrow 0) \,.
\label{Eq:D_operator}
\end{equation}

Among the three terms composing $\hat{U}(t_f\! \leftarrow\! 0)$ in Eq.\;(\ref{Eq:Dyson_Eq_3}), $\hat{U}_d(t_f)$ is responsible for direct ionization whereas $\hat{U}_r(t_f)$ includes recollision. It is to be stressed that Eq.\,(\ref{Eq:Dyson_Eq_3}), with the definitions given in Eqs.\,(\ref{Eq:Evolution_Operators}) and\,(\ref{Eq:D_operator}), is still exact. This type of Dyson expansion could be iterated by considering multiple ionization and higher order recollisions. In the present simplified model we stop at this second order decomposition.

Now, using these equations, we can split the ionization amplitude in two contributions:
\begin{equation}
a(k_x,k_y) = a_d(k_x,k_y) + a_r(k_x,k_y)\,,
\label{Eq:Analytic_tran_amp_2}
\end{equation}
with
\begin{equation}
a_d(k_x,k_y) = \langle \Psi^{+}_{\bm{k}}|\,\hat{U}_d(t_f)\,|\Psi(0)\rangle
\label{Eq:direct_amp_SFA1}
\end{equation}
and
\begin{equation}
a_r(k_x,k_y) = \langle \Psi^{+}_{\bm{k}}|\,\hat{U}_r(t_f)\,|\Psi(0)\rangle\,.
\label{Eq:Recollision_amp_SFA1}
\end{equation}
Eq.\,(\ref{Eq:direct_amp_SFA1}) gives the transition amplitude associated with direct ionization whilst Eq.\,(\ref{Eq:Recollision_amp_SFA1}) gives  the transition amplitude associated with ionization preceded by recollision. Hence, the 2D LIED spectrum can be written as
\begin{equation}
\mathcal{I}(k_x,k_y) = \big|a_d(k_x,k_y) + a_r(k_x,k_y)\big|^2\,,
\label{Eq:Trans_prob_1}
\end{equation}
an expression which shows clearly the appearance of an interference between the direct and recolliding ionization pathways. Note that such expression is common when describing strong field ionization using a SFA approach\;\cite{Suarez_2015}.

%%%%%%%%%%%%%%%%%%%%%%%%%%%%%%%%%%%%%%%%%%%%%%%%%
\subsubsection{Approximate Transition Amplitude}%
%%%%%%%%%%%%%%%%%%%%%%%%%%%%%%%%%%%%%%%%%%%%%%%%%

Evaluating the direct ionization amplitude $a_d(k_x,k_y)$ is relatively easy as compared to the recollision amplitude $a_r(k_x,k_y)$ because of the appearance of \mbox{$\hat{U}(t_f \!\leftarrow \!t'')$} in the expression of $\hat{U}_r(t_f)$. To make this evaluation tractable, we use the Strong Field Approximation (SFA)\;\cite{PhysRevA.74.063404, PhysRevA.78.033412}, and we replace $\hat{U}(t_f\!\leftarrow\!t'')$ by the Volkov evolution operator $\hat{U}_v(t_f\!\leftarrow\!t'')$, with
\begin{equation}
\hat{U}_r(t_f) \simeq \int_{0}^{t_f}\!\!\!\!dt'\!\!\int_{t'}^{t_f}\!\!\!\!dt''\,\hat{U}_v(t_f\! \leftarrow\! t'')\,V(\bm{r})\,\mathcal{D}(t'',t')\,.
\label{Eq:ApproxUr}
\end{equation}

Replacing $\hat{U}(t_f\!\leftarrow\!t'')$ by $\hat{U}_v(t_f\!\leftarrow\!t'')$ in Eq.\,(\ref{Eq:ApproxUr}) means that after the first recollision event, we neglect the Coulomb force compared to the interacting IR field, an approximation valid in the asymptotic region, where the Coulomb interaction is negligible.

As a second step for simplifying the model, the outgoing waves $|\Psi^{+}_{\bm{k}}\rangle$ are approximated by plane waves $|\Phi^{\text{pw}}_{\bm{k}}\rangle$. This approximation is justified asymptotically. Within these approximations we obtain
\begin{equation}
a_d(k_x,k_y) \simeq\, i \int_{0}^{t_f}\!\!dt'\,e^{-i\bar{S}_1}\,
 \big\langle\Phi^{\text{pw}}_{\bm{k'}}\,\big|\,\hat{\bm{\mu}}\cdot\bm{E}(t')\,\big|\Psi(0)\big\rangle\,,
\label{Eq:direct_amp_SFA3}
\end{equation}
and
\begin{eqnarray}
a_r(k_x,k_y) & \simeq & \int_{0}^{t_f}\!\!E(t')dt'\!\int_{t'}^{t_f}\!\!dt''\, e^{-i\bar{S}_2}\nonumber\\
             &        & \big\langle\Phi^{\text{pw}}_{\bm{k''}}\,\big|\,V(\bm{r})\,
                        \big|\Psi_r\big\rangle\,,
\label{Eq:Recollision_amp_SFA3}
\end{eqnarray}
where
\begin{subequations}
\begin{eqnarray}
\bar{S}_1 & = & \frac{1}{2}\int_{t'}^{t_f}  \left[\bm{k}+\bm{A}(\tau)\right]^2d\tau - I_p\,t'\,,
\label{Eq:Sbar_1} \\
\bm{k'}   & = & \bm{k}+\bm{A}(t')\,,
\label{Eq:kprime}\\
\bar{S}_2 & = & \frac{1}{2}\int_{t''}^{t_f} \left[\bm{k}+\bm{A}(\tau)\right]^2d\tau - I_p\,t'\,,
\label{Eq:Sbar_2}  \\
\bm{k''}  & = & \bm{k}+\bm{A}(t'')\,,
\label{Eq:ksecond} \\
\big|\Psi_r\big\rangle & = & \hat{U}_v(t''\!\leftarrow\!t')\,x\,\big|\Psi(0)\big\rangle\,.
\end{eqnarray}
\end{subequations}

In Section\;\ref{Sec:Num_Res}, it has been noted that the most interesting features of the photoelectron spectrum lie in the high momentum ($k_y$) part of the 1D averaged spectra. This is because these electrons are characterized by de Broglie wavelengths small enough to resolve sub-\AA\/ spatial scales. Thus, describing accurately the low-energy part of the spectrum and the parallel momentum ($k_x$) distribution of the photoelectrons is not essential. It has also been shown that large energies are reached by electrons ionized around a maximum of the field and hence around a minimum of the potential vector\;\cite{PhysRevLett.71.1994, PhysRevA.54.742}. Thus $\bm{A}(t')$ can be neglected in Eq.\,(\ref{Eq:kprime}). In addition electrons with high kinetic energies mainly recollide with the ionic core at a minimum of the field, corresponding to a maximum of the vector potential\;\cite{PhysRevLett.71.1994, PhysRevA.54.742}. In Eq.\,(\ref{Eq:ksecond}) we will therefore use $\bm{A}(t'') \simeq \pm E_0/\omega_L\,\hat{\bm{x}}$. The potential vector $\bm{A}(t'')$ therefore induces a strong shift on the parallel component $k_x$ of the electron momentum. In practice, this shift is of no significance in the present approach, since it will be averaged out in the calculation of Eq.\,(\ref{Eq:1D_spec_Eqn}) and we will therefore not take it into account in the following.

Within these approximations and to avoid discrepancies between the SFA spectrum and the spectrum obtained from the solution of the TDSE, one should restrict the analysis of the 1D averaged signal to the highest $k_y$ momentum components only. The interest of this severe approximation lies however in the fact that it simplifies the model by allowing the separation of the temporal from the spatial integrals involved in Eq.\,(\ref{Eq:direct_amp_SFA3}). Thus, for the direct ionization amplitudes, one has:
\begin{equation}
a_d(k_x,k_y) \simeq A_{d}\,\langle\Phi^{\text{pw}}_{\bm{k}}|\,x\,|\Psi(0)\rangle\,,
\label{Eq:Appox_ad}
\end{equation}
where
\begin{equation}
A_{d} = i\int_{0}^{t_f} E(t')\,e^{-i\bar{S}_1}\,dt' \,.
\end{equation}
Similarly, the recollision amplitude becomes
\begin{eqnarray}
a_r(k_x,k_y) & \simeq & \int_{0}^{t_f}\!\!E(t')dt'\!\int_{t'}^{t_f}\!\!dt''\,e^{-i\bar{S}_2}\,\nonumber\\
             &        & \big\langle\Phi^{\text{pw}}_{\bm{k}}\,\big|\,V(\bm{r})\,\big|\Psi_r\big\rangle\,.
\label{Eq:Appox_ar}
\end{eqnarray}
Using the closure property of the plane wave basis set one obtains
\begin{equation}
\big|\Psi_r\big\rangle = \int\!\!d\bm{k'} e^{-ik'^2\Delta t/2} \big\langle\Phi^{\text{pw}}_{\bm{k'}}\big|x\big|\Psi(0)\big\rangle\,
\big|\Phi^{\text{pw}}_{\bm{k'}}\big\rangle\,,
\end{equation}
where $\Delta t = t'' - t' \simeq 0.7\,(2\pi/\omega_L)$ is the mean time during which the electron wave packet propagates in the continuum\;\cite{PhysRevLett.71.1994, PhysRevA.54.742}. The temporal and spatial integrals can thus be separated in the expression (\ref{Eq:Appox_ar}) of the recollision amplitude as
\begin{equation}
a_r(k_x,k_y) \simeq A_{r}\,\big\langle\Phi^{\text{pw}}_{\bm{k}}\big|\,V(\bm{r})\,\big|\Psi_r\big\rangle\,,
\label{Eq:ar_k_Approx}
\end{equation}
where
\begin{equation}
A_{r} = \int_{0}^{t_f}\!\!dt'\int_{t'}^{t_f}\!\!dt''\,E(t')\,e^{-i\bar{S}_2}\,.
\label{Eq:Const_amp_recoll}
\end{equation}

Finally, the approximate transition amplitude is given by
\begin{equation}
a(k_x,k_y) \simeq A_{d}\big\langle\Phi^{\text{pw}}_{\bm{k}}\big|x        \big|\Psi(0)\big\rangle +
                  A_{r}\big\langle\Phi^{\text{pw}}_{\bm{k}}\big|V(\bm{r})\big|\Psi_r \big\rangle\,.
\label{Eq:Trans_amp_approximate_general}
\end{equation}
This equation has to be developed on a suitable basis of initial states for the final analytical form of the LIED spectra.

%%%%%%%%%%%%%%%%%%%%%%%%%%%%%%%%%%%%%%%%%%%%%
\subsection{Initial Molecular Wave Function}%
\label{subsec:ini_guess}                    %
%%%%%%%%%%%%%%%%%%%%%%%%%%%%%%%%%%%%%%%%%%%%%

Eq.\,(\ref{Eq:Trans_amp_approximate_general}) expresses the ionization amplitude as a sum two terms, each written in the form  of a product of spatial and temporal integrals.The first term is associated with direct ionization and, the second one with recollision events. As we see from the LIED spectra, the information we are interested in is encoded in the spatial integrals. Thus in the following discussions, the temporal integrals $A_{d}$ and $A_{r}$ will be taken as adjustable coefficients in order to match the approximate spectrum with the calculated spectrum.

To proceed further with the evaluation of the spatial integrals, we need to specify the initial wave function $\Psi(\bm{r},0)=\big\langle\bm{r}\big|\Psi(0)\big\rangle$. In so far as SAE is valid, this is a molecular orbital. In quantum chemistry, this is usually expressed as a linear combination of atomic orbitals (LCAO method) and there are many basis set ansatz for representing localized atomic wave functions. Here the initial HOMO orbital is taken as an anti-symmetric linear combination of $2p_x$ atomic orbitals (see Fig.\;\ref{fig:system}(b))
\begin{subequations}
\begin{eqnarray}
\Psi(\bm{r},0) & = & \Phi_{2p_x}(\bm{r}+\bm{R}) - \Phi_{2p_x}(\bm{r}-\bm{R})\\
               & = & \Phi_{2p_x}^{-}(\bm{r})    - \Phi_{2p_x}^{+}(\bm{r})\,.
\label{Eq:Ini_Guess_2px}
\end{eqnarray}
\end{subequations}
Ideally one would choose for $\Phi_{2p_x}(\bm{r})$ a Slater-type orbital of the form
\begin{eqnarray}
\Phi_{2p_x}^s(\bm{r}) &=& \mathcal{N}_s\;x\;\mathrm{e}^{-\zeta\,r}\, , \quad r=(x^2+y^2)^{1/2}
 %               %      &=& \mathcal{N}_s\;r\cos\theta_r\;\mathrm{e}^{-\zeta\,r}\,,
\label{Eq:STO_2px}
\end{eqnarray}
with the normalization factor \mbox{$\mathcal{N}_s = \zeta^2 \sqrt{8/3\pi}$} in two dimensions, where $\zeta$ is the Slater exponent. This analytical form, once introduced in Eq.\,(\ref{Eq:Ini_Guess_2px}), is a reasonable candidate for representing the HOMO orbital but an important disadvantage then lies in the difficulty of evaluating multi-center integrals such as the recolliding integral of Eq.\,(\ref{Eq:Trans_amp_approximate_general}). It appears that this difficulty can be removed if the Slater orbital (\ref{Eq:STO_2px}) is replaced by a Gaussian type orbital of the form
\begin{equation}
\Phi_{2p_x}^g(\bm{r}) = \mathcal{N}_g\;r\cos\theta_r\;\mathrm{e}^{-\alpha\,r^2}\,,
\label{Eq:STO-1G-2px}
\end{equation}
with the normalization factor \mbox{$ \mathcal{N}_g = \alpha\,\sqrt{8/\pi}$}, where $\alpha$ is the Gaussian exponent. Actually, the Gaussian function (\ref{Eq:STO-1G-2px}) can be made a good approximation of the Slater orbital\;(\ref{Eq:STO_2px}) with an appropriate choice of $\alpha$.

\begin{figure}[ht]
\includegraphics[width=8.5cm]{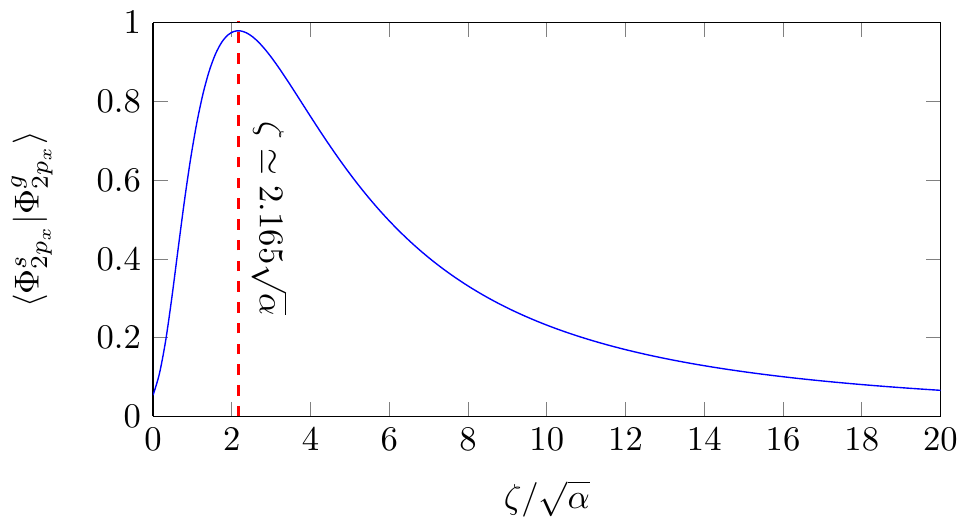}
\caption{(Color online). Overlap between the Gaussian-type (\ref{Eq:STO-1G-2px}) and Slater-type (\ref{Eq:STO_2px}) orbitals used in the present study as a function of the dimensionless ratio $\zeta/\sqrt{\alpha}$ (see text for details).}
\label{fig:STO-GTO-overlap}
\end{figure}

Fig.\;\ref{fig:STO-GTO-overlap} shows the overlap between the wave functions (\ref{Eq:STO-1G-2px}) and (\ref{Eq:STO_2px}) as a function of the dimensionless ratio $\zeta/\sqrt{\alpha}$. It is clear that for $\zeta \simeq 2.165\,\sqrt{\alpha}$, the Slater and Gaussian orbitals are very similar, with an overlap of about $98\%$. In the following, for the evaluation of the integrals, we will use Gaussian-type atomic orbitals, but for the reconstruction of the initial molecular state we will use Slater-type orbitals whose size are defined from the optimal ratio $\zeta/\sqrt{\alpha} = 2.165$.

%%%%%%%%%%%%%%%%%%%%%%%%%%%%%%%%%%%%%%%%%%%%%%%%%%%
\subsection{Approximate 1D Photoelectron Spectrum}%
%%%%%%%%%%%%%%%%%%%%%%%%%%%%%%%%%%%%%%%%%%%%%%%%%%%

The expression for the approximate transition amplitude $a(k_x,k_y)$ given in Eq.\,(\ref{Eq:Trans_amp_approximate_general}) can now be evaluated for the initial HOMO wave function given in Eq.\,(\ref{Eq:Ini_Guess_2px}), as
\begin{eqnarray}
a(k_x,k_y) & = & A_{d}\Big[\langle\Phi^{\text{pw}}_{\bm{k}}|\,x\,|\Phi_{\mathrm{2p}_x}^{-}\rangle
                         - \langle\Phi^{\text{pw}}_{\bm{k}}|\,x\,|\Phi_{\mathrm{2p}_x}^{+}\rangle\Big]
\nonumber\\
           & + & A_{r}\Big[\langle\Phi^{\text{pw}}_{\bm{k}}|\,V\,|\Phi_{\text{rec}}^{-}\rangle
                         - \langle\Phi^{\text{pw}}_{\bm{k}}|\,V\,|\Phi_{\text{rec}}^{+}\rangle\Big]
\label{Eq:Trans_amp_CO2_general1}
\end{eqnarray}
where $|\Phi_{\text{rec}}^{\pm}\rangle$ denotes
\begin{equation}
|\Phi_{\text{rec}}^{\pm}\rangle = \int\!d\bm{k'} e^{-ik'^2\Delta t/2}
\langle\Phi^{\text{pw}}_{\bm{k'}}|\,x\,|\Phi_{\mathrm{2p}_x}^{\pm}\rangle\;
|\Phi^{\text{pw}}_{\bm{k'}}\rangle\label{Phi_rec_pm}\,.
\end{equation}

The first two integrals in Eq.\,(\ref{Eq:Trans_amp_CO2_general1}) represent direct ionization from displaced (oxygen $2p_x$) orbitals and the last two represent the ionization amplitudes after a recollision event. The two integrals associated with direct ionization amplitudes are just Fourier transforms (FT) of products of the dipole operator $x$ with displaced $2p_x$ orbitals. In momentum space, this spatial translation becomes a simple phase shift of the form $\exp\left[\pm ik_{y}R\right]$ of the FT signal of $\Phi_{\mathrm{2p}_x}^{\pm}$. Taking this simplification into account, Eq.\,(\ref{Eq:Trans_amp_CO2_general1}) can be reduced to
\begin{eqnarray}
a(k_x,k_y) & = & A_{d}\,\sin(k_yR)\,\langle\Phi^{\text{pw}}_{\bm{k}}|\,x\,|\Phi_{\mathrm{2p}_x}\rangle\nonumber\\
           & + & A_{r}\Big[\langle\Phi^{\text{pw}}_{\bm{k}}|\,V\,|\Phi_{\text{rec}}^{-}\rangle
                         - \langle\Phi^{\text{pw}}_{\bm{k}}|\,V\,|\Phi_{\text{rec}}^{+}\rangle\Big]
\label{Eq:Trans_amp_CO2_general2}
\end{eqnarray}

The evaluation of the direct ionization amplitude using a Gaussian-type orbital yields
\begin{equation}
 a_{d}(k_x,k_y) =  A_{d}\,\sin(k_yR)\,(k_x^2-2\alpha)\,e^{-\frac{k_x^2+k_y^2}{4\alpha}}\,,
\label{Eq:direct_CO2_g}
\end{equation}
provided $A_{d}$ accounts for all constant factors.

The calculation of the recollision amplitude $a_{r}(k_x,k_y)$ is more involved since it needs the knowledge of the functional form of the recolliding wave functions $\Phi_{\text{rec}}^{\pm}(\vec r)$. Using for the initial state a Gaussian-type orbital $\Phi_{\mathrm{2p}_x}$ located at origin we obtain
\begin{equation}
\Phi_{\text{rec}}^{0}(r) \propto \frac{\alpha-i\beta-2\beta^2x^2}{(\alpha-i\beta)^{3}}\;e^{i\gamma r^2}\,,
\label{Eq:Convolution_Carbon}
\end{equation}
where $\gamma=\alpha\beta/(\alpha-i\beta)$ and $\beta=1/(2\Delta t)$. The wave functions $\Phi_{\text{rec}}^{\pm}(r)$ are identical to $\Phi_{\text{rec}}^{0}(r)$ except for a phase shift, so that the corresponding recollision wave functions are given by
\begin{subequations}
\begin{eqnarray}
\Phi_{\text{rec}}^{-}(r) & = & e^{i\gamma R^2}\,e^{+i2y\gamma R}\;\Phi_{\text{rec}}^{0}(r)\,,\\
\Phi_{\text{rec}}^{+}(r) & = & e^{i\gamma R^2}\,e^{-i2y\gamma R}\;\Phi_{\text{rec}}^{0}(r)\,.
\end{eqnarray}
\label{Eq:Spatial_Covolution_CO2}
\end{subequations}
In the near IR ($\lambda$ = 800\,nm to 2.5\,$\mu$m) the parameter $\beta$ of Eq.\,(\ref{Eq:Convolution_Carbon}) is in the range $10^{-2}$ to $10^{-3}$\;a.u. In comparison, the Gaussian orbital exponent $\alpha$ is usually of the order of 1\,a.u. These orders of magnitudes can be used in order to simplify further the expression of the ionization amplitude.

Since the binding potential $V(\bm{r})$ is characterized by three attractive centers, the recollision amplitude $a_{r}(k_x,k_y)$ (second part on the \textit{r.h.s} of Eq.\,(\ref{Eq:Trans_amp_CO2_general2})) contains, for the HOMO of CO$_2$, 6 integrals. Indeed, from the HOMO, ionization may originate from any of the two oxygen atoms and recollision may take place on any of the three atoms. Fortunately, these 6 integrals are similar. In the case of the HOMO, the electron wave packet is launched from both of the oxygen atoms marked as O$_1$ and O$_2$ in Fig.\;\ref{fig:system}. On recollision, the contribution from the first oxygen atom O$_1$ will scatter from the parent atom O$_1$ itself  as well as from the two neighboring atoms: from the carbon atom C and  from the second oxygen atom O$_2$. This part of the rescattering amplitude, shown in Fig.\;\ref{fig:system}, can be written as
\begin{equation}
a_{r}^{\text{O}_1}(k_x,k_y) = \langle\Phi^{\text{pw}}_{\bm{k}}|\,V\,|\Phi^{+}_{\text{rec}}\rangle\,.
\end{equation}
where the three-center potential $V$, supposed to be of a Coulomb form, is given by
\begin{equation}
V(\bm{r}) = - \dfrac{q_{\mathrm{O}}}{|\bm{r}+\bm{R}|} - \dfrac{q_{\mathrm{C}}}{|\bm{r}|}
            - \dfrac{q_{\mathrm{O}}}{|\bm{r}-\bm{R}|}\,.
\label{Eq:SoftCoulombPotential_SAE-molecule-1bis}
\end{equation}
What matters most for the recollision is the scattering taking place in the vicinity of Coulombic cores. At first order near the singularities of the potential wells, \textit{i.e.} for $x \rightarrow 0$ and $y \rightarrow \{\,-R,\,0,\,R\,\}$, and taking into account Eqs.\,(\ref{Eq:Convolution_Carbon}) and\;(\ref{Eq:Spatial_Covolution_CO2}), the above integral can be reduced to
\begin{equation}
a_{r}^{\text{O}_1}(k_x,k_y) \propto
\frac{-\mathrm{e}^{ik_y R}-\mathrm{e}^{i\beta R^{2}}-\mathrm{e}^{-ik_y R}\,\mathrm{e}^{i\beta 4R^{2}}}{|k_y|}\,.
\label{Eq:Approx_recoll_1st_O}
\end{equation}
Similarly, for the wave packet originating from the second oxygen atom, we obtain:
\begin{eqnarray}
a_{r}^{\text{O}_2}(k_x,k_y) & = &
  \langle\Phi^{\text{pw}}_{\bm{k}}|\,V\,|\Phi^{-}_{\text{rec}}\rangle\,\notag\\
  & \propto &
  \frac{\mathrm{e}^{-ik_y R}+\mathrm{e}^{i\beta R^{2}}+\mathrm{e}^{ik_y R}\,\mathrm{e}^{i\beta 4R^{2}}}{|k_y|}\,.
  \label{Eq:Approx_recoll_2nd_O}
\end{eqnarray}

Finally, the total recollision amplitude is
\begin{equation}
a_{r}(k_x,k_y) = A_{r}\,\frac{1-\mathrm{e}^{i4\beta R^{2}}}{|k_{y}|}\,\sin(k_{y}R)\,.
\label{Eq:HOMO_recollision_CO2}
\end{equation}
Combining Eqs.\,(\ref{Eq:direct_CO2_g}) and\;(\ref{Eq:HOMO_recollision_CO2}) we obtain the 2D transition amplitude. The transition probability is the square modulus of this transition amplitude. Finally, averaging over the parallel momentum component $k_x$, the 1D spectrum is written as
\begin{equation}
\mathcal{S}(k_y)=\left(|A_d|^2\,\mathrm{e}^{-\frac{k_{y}^2}{2\alpha}}+ \frac{|A_r|^2}{k_y^2}\,\right)\, \sin^2(k_{y}R)\,.
\label{Eq:1D_spectrum_HOMO}
\end{equation}
This is the compact analytical form we will use in the next section for our inversion procedure.

%%%%%%%%%%%%%%%%%%%%%%%%%%%%%%%%%%%%%%%%%%%%%%
\section{Results: Reconstruction of Orbitals}%
\label{Sec:Reconstruction}                   %
%%%%%%%%%%%%%%%%%%%%%%%%%%%%%%%%%%%%%%%%%%%%%%

Eq.\,(\ref{Eq:1D_spectrum_HOMO}) is the final result we intended to derive for solving the inverse problem. Taking $|A_d|$, $|A_r|$, $\alpha$ and $R$ as four independent adjustment variables, this expression can be compared with 1D averaged LIED spectra calculated from the solution of the TDSE. In general, the model can be used for any internuclear distances of the CO$_2$ molecule. However, as discussed previously, with the particular laser parameters chosen in the present study our model is not expected to perform well for small values of $R$. We thus chose only two cases for this comparison: $R=3.5$\,\AA\/ and $R=5.0$\,\AA\/.

To ease the multi-parameter fitting procedure, it is well-known that the search for the best fit should start from a good guess value. Here the range of the parameter $R$ can be obtained easily from the spectrum itself by measuring the fringe width $\Delta k = \pi/R$ as discussed in\;\cite{PhysRevA.85.053417}. Thus we are left with three completely unknown parameters and one partially known parameter. The fitting process is performed here by using the well-known Levenberg-Marquardt algorithm (LMA)\;\cite{Algorithm}, because of its robustness to find the best possible solutions even if the procedure starts with initial guess values relatively far from the final one.

\begin{table}[ht]
	\caption{Fitted values of the parameters involved in the SFA analytical model of Eq.\,(\ref{Eq:1D_spectrum_HOMO}).}
	\begin{ruledtabular}
		\begin{tabular}{c c c c c}
			                 & $A_d$ (au) & $A_r$ (au) & \;\;\;\;$R$ (\AA)\;\; & $\alpha$ (au) \\ [0.5ex]
			\hline \noalign{\vskip 1.0ex}
			                 & 0.00478    & 0.000808   & 3.628                 & 0.535 \\ [0.5ex]
            For $R=3.5$\,\AA & 0.00506    & 0.000720   & 3.616                 & 0.516 \\ [0.5ex]
			                 & 0.00513    & 0.000634   & 3.624                 & 0.527 \\ [0.5ex]
			                 & 0.00405    & 0.001087   & 3.619                 & 0.520 \\ [0.5ex]
			\hline \noalign{\vskip 1.0ex}
			Average          & 0.00476    & 0.000812   & 3.622                 & 0.525 \\ [0.5ex]
			\hline \noalign{\vskip 0.5ex}
			\hline \noalign{\vskip 1.0ex}
		     	             & 0.0157    & 0.00385   & 5.141                 & 0.676 \\ [0.5ex]
			For $R=5.0$\,\AA & 0.0121    & 0.00639   & 5.141                 & 0.625 \\ [0.5ex]
			 	             & 0.0137    & 0.00507   & 5.142                 & 0.657 \\ [0.5ex]
			 	             & 0.0145    & 0.00452   & 5.142                 & 0.669 \\ [0.5ex]
			\hline \noalign{\vskip 1.0ex}
			Average          & 0.0140    & 0.00496   & 5.142                 & 0.657 \\ [0.5ex]
		\end{tabular}
	\end{ruledtabular}
	\label{Table_fit}
\end{table}

The fitting process is performed on the high kinetic energy part of the spectra. The highest accessible kinetic energy and hence the highest momentum component $k_y^{max}$ is defined by the cut-off energy $3.17\,U_p$, which is fixed by the laser parameters used in the calculation or experiment. In order to obtain reliable values for the parameters, the fitting process must be repeated several times. This is done by varying the lower limit of the kinetic momentum $k_y^{min}$ taken into account, between $1.15$\,au and $1.25$\,au in the present calculation.

Values of the relevant parameters obtained for four different lower limits are given in table\;\ref{Table_fit}. The values obtained for the internuclear distance $R$ are very stable and accurate. In addition, the values obtained for the orbital exponent $\alpha$ are also relatively stable. Since these two parameters are the ingredients used to reconstruct the molecular orbital (\ref{Eq:Ini_Guess_2px}) they will lead to a very similar orbital, whatever the other parameters chosen in Table\;\ref{Table_fit}. In practice, we average over several fits in order to extract these parameters (see Table\;\ref{Table_fit}).

\begin{figure}[ht]
\includegraphics[width=7.5cm]{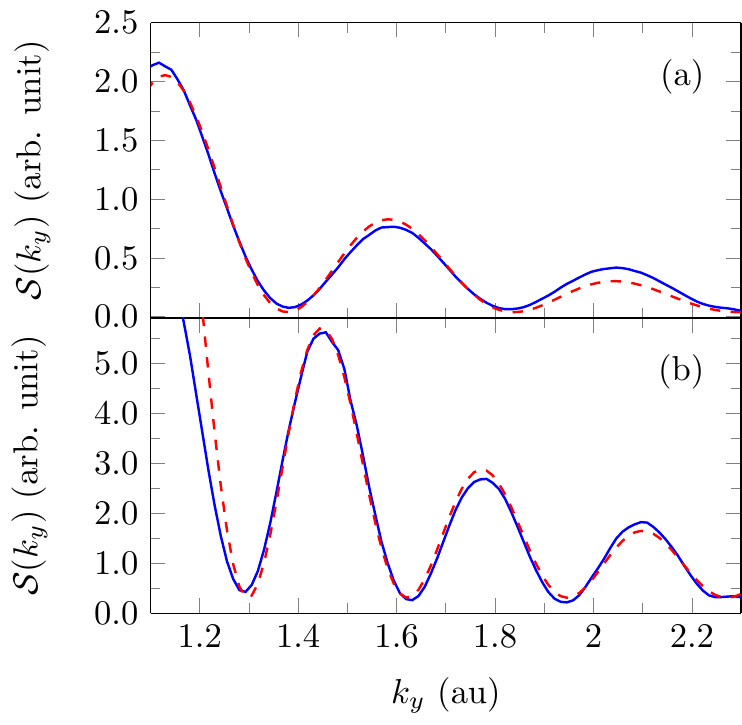}
\caption{(Color online). 1D averaged LIED spectra $\mathcal{S}(k_y)$. The blue solid lines are the spectra calculated using the time-dependent Schr\"odinger equation and the red dashed lines show the results of the best fits using the analytical SFA model. Panel (a) is for $R=3.5\,$\AA\/ and panel (b) is for $R=5.0\,$\AA.}
\label{fig:Fit_spectra}
\end{figure}

Typical numerical and model 1D LIED spectra are given in Fig.\;\ref{fig:Fit_spectra}. Panel (a) is for $R=3.5$\,\AA\/ and panel (b) is for $R=5.0$\,\AA. The numerical spectra $\mathcal{S}(k_y)$ obtained by solving the TDSE are shown as bold blue curves and the fitted (model) spectra are shown as red dashed curves, as a function of $k_y$. As it is apparent from the figure, the model and the numerical calculations fit well. For both cases considered here, the relative errors in the retrieved internuclear distances are of the order of $3\,\%\,$: We obtained $R=3.62\,$\AA\/ instead of $3.50\,$\AA\/ and $R=5.14\,$\AA\/ instead of $5.00\,$\AA\/ (see averaged $R$ values in table\;\ref{Table_fit}).

\begin{figure}[ht]
\includegraphics[width=7.5cm]{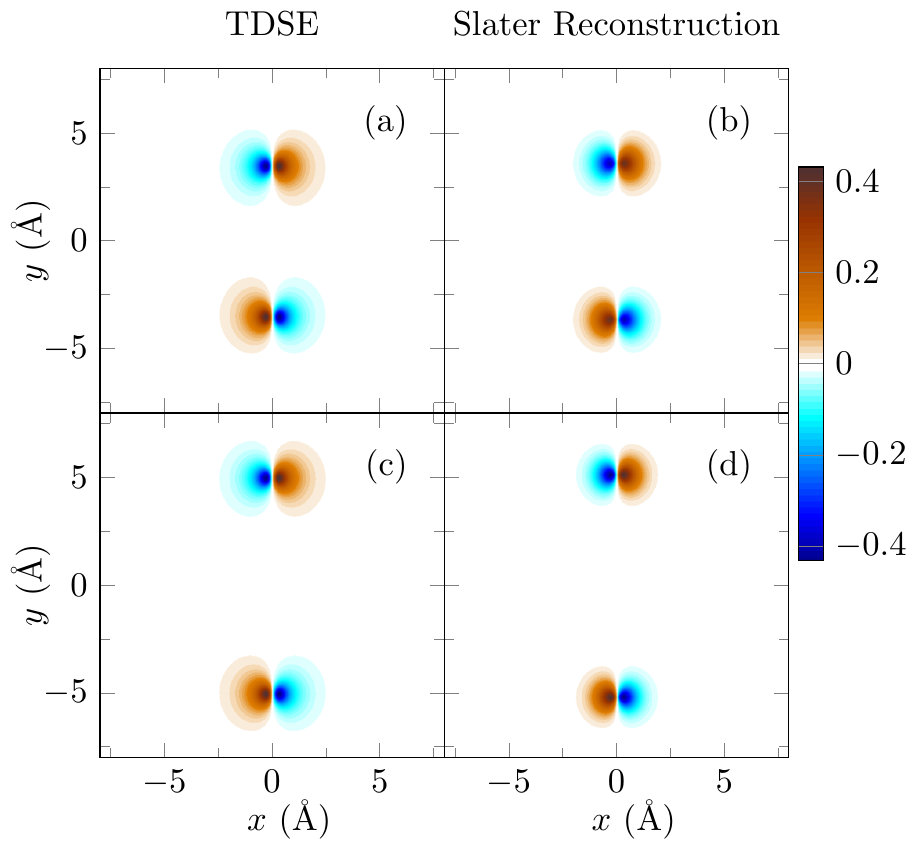}
\caption{(Color online). Panels (a) and (c): Initial wave functions used for the TDSE calculation with $R=3.5$\,\AA\/ and $5.0$\,\AA\/ respectively. Panels (b) and (d): Associated reconstructed molecular orbitals.}
\label{fig:Recontruct}
\end{figure}

Taking the average values of the fitted internuclear distances $R$ and Gaussian exponents $\alpha$ we can reconstruct the initial state used for deriving Eq.\,(\ref{Eq:1D_spectrum_HOMO}). Finally, as discussed in Section\;\ref{subsec:ini_guess}, this initial state can also be given in terms of Slater-type orbitals. These functions will give the best possible simple form of the initial state. Reconstructed approximate Slater forms of the initial states are given in Fig.\,\ref{fig:Recontruct}. Panels (a) and (c) are the initial states used in the TDSE calculation for $R=3.5$\,\AA\/ and $R=5.0$\,\AA. Panels (b) and (d) are the corresponding reconstructed molecular orbitals.

The overlap between the reconstructed orbital and the initial state used in the numerical TDSE calculations is higher than $96\%$: $96.3\%$ for $R=3.5\,$\AA\/ and $97.2\%$ for $R=5.0\,$\AA. This reconstruction shows that for large internuclear distances LIED techniques could be used to image molecular orbitals with a rather good accuracy using a simple multi-parameter fitting procedure.

\begin{figure}[ht]
\includegraphics[width=8.5cm]{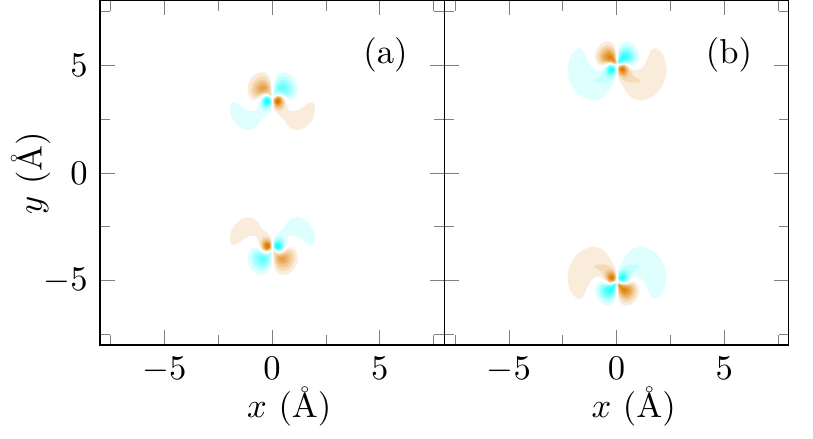}
\caption{(Color online). Difference between the initial states and their reconstructions. Panel (a) is for $R=3.5\,$\AA\/ and panel (b) is for $R=5.0\,$\AA. The color map used is the same as in Fig.\;\ref{fig:Recontruct}.}
\label{fig:Error}
\end{figure}

It is also possible to depict the discrepancy in the reconstructed orbitals caused by the inaccuracies in $R$ and $\alpha$ by plotting the difference between the exact and reconstructed orbitals. These differences are shown in Fig.\;\ref{fig:Error} using the same color code as in Fig.\;\ref{fig:Recontruct}. Panel (a) is for  $R=3.5\,$\AA\/ and panel (b) is for $R=5.0\,$\AA. The discrepancy shown in this figure is due to the combined errors in the reconstructed values of both the orbital exponent $\alpha$ and the internuclear distance $R$.

Any error in the internuclear distance $R$ would be crucial since it would cause a significant mismatch in the location of the reconstructed orbital. Here, since the fitted value of $R$ is very close to its exact value, this problem does not appear.

A small error in the orbital exponent $\alpha$ is, on the other hand, not as crucial since the overlap between the exact wave function and the reconstructed orbital varies smoothly with $\alpha$. We have calculated the optimal $\alpha$ values for our exact initial states by computing the overlap between the initial state and the LCAO form we have adopted in this study. We have obtained $\alpha_{\mathrm{opt}} = 0.624\,$au for $R=3.5\,$\AA\/ and $\alpha_{\mathrm{opt}} = 0.626\,$au for $R=5.0\,$\AA. The relative errors in the fitted values of $\alpha$ are therefore of the order of 16\% for $R=3.5\,$\AA\/ and 5\% for $R=5.0\,$\AA. We again see here that our inversion procedure is more accurate for the largest internuclear distance.

Being a model developed using a single active electron approximation, analyzing LIED processes with multi-electron ionization channels may give additional discrepancies in the retrieved values of the parameters. One of the main problem for including interactions between electrons is the difficulty to solve such situations analytically. Aiming for a compact analytical form given with a relatively small number of fitting parameters, an extension of the model beyond the single active electron approximation is far from trivial in the spirit of an inverse problem.

Other approximations could be relaxed more easily. For example, higher process of the recollision events can be included into the picture by extending the strong field approximation to the desired higher order terms. This may improve the model, but to our best understanding, the second order development used here retains the main elements necessary for an accurate reconstruction procedure for linear molecules with large internuclear distances ($R > 3\,$\AA\/).

The inverse problem discussed in this manuscript in the case of the HOMO orbital of the CO$_2$ molecule can be relatively easily extended to the deeper \mbox{HOMO-1} orbital by modifying slightly the analytical model. In this case, the atomic orbitals of the three composite atoms have a significant overlap and form a symmetric molecular orbital. But the relative contributions of the C and O atoms are different. This gives an additional parameter which should be introduced in the model. This additional parameter would also have to be retrieved by an inversion procedure. It should also be relatively easy to make some other simple modifications to the analytical model to treat other linear molecules.

%%%%%%%%%%%%%%%%%%%%%
\section{Conclusion}%
\label{Sec:Conclude}%
%%%%%%%%%%%%%%%%%%%%%

In this paper, we discuss some possibilities for imaging molecular orbitals offered by laser induced electron diffraction following the strong field ionization of a pre-aligned linear molecule. The problem is discussed in detail for the HOMO orbital of the carbon dioxide molecule.

The system is described theoretically in the framework of a single active electron model. The strong field photoelectron spectra are obtained by solving the time-dependent Schr\"odinger equation (TDSE) for different initial internuclear distances. An approximate, but compact analytical model is developed for these photoelectron spectra using three classes of approximations: (i)\;using the single active electron approximation, (ii)\;using the strong field approximation and (iii)\;using an approximate LCAO ansatz for the initial molecular orbital.

This analytical model contains some parameters which are fitted by comparison with the TDSE results. This fitting procedure allows for the extraction of the internuclear distance and the corresponding Slater-type orbital exponents. The initial ansatz for the molecular orbital is then reconstructed with these parameters, providing an accurate representation of the initial state used in the TDSE, with an overlap which is higher than $96\%$.

This approach can be effectively used for the reconstruction of the HOMO molecular orbital with a good accuracy. It should be possible to extend this model to other initial orbitals and to other linear molecules. In the future, the inclusion of the nuclear dynamics could enable this model to image reaction dynamics like the photo-dissociation of linear molecules, for instance.

%%%%%%%%%%%%%%%%%%%%%%%%%%
\section*{Acknowledgment}%
%%%%%%%%%%%%%%%%%%%%%%%%%%

We thank Misha Ivanov for stimulating discussions. R.P.J. and E.C. acknowledge support from  the EU (Project ITN-2010-264951, CORINF). We also acknowledge the use of the computing center GMPCS of the LUMAT federation (FR LUMAT 2764).

%%%%%%%%%%%%%%%%%%%%%%%%%%%%
%
%%%%%%%%%%%%%%%%%%%%%%


\begin{thebibliography}{99}%
%%%%%%%%%%%%%%%%%%%%%%%%%%%%

\bibitem{PhysRevLett.114.255501}
M. P. Minitti, J. M. Budarz, A. Kirrander, J. S. Robinson, D. Ratner, T. J. Lane, D. Zhu, J. M. Glownia, M. Kozina, H. T. Lemke, M. Sikorski, Y. Feng, S. Nelson, K. Saita, B. Stankus, T. Northey, J. B. Hastings, and P. M. Weber, \textit{Phys. Rev. Lett.} \textbf{114}, 255501 (2015).

\bibitem{PhysRevLett.109.133202}
C. J. Hensley, J. Yang, and Martin Centurion, \textit{Phys. Rev. Lett.} \textbf{109}, 133202 (2012).

\bibitem{JPhysB.49.062001}
F. Calegari, G. Sansone, S. Stagira, C. Vozzi, and M. Nisoli, \textit{J. Phys. B - At. Mol. Opt. Phys.} \textbf{49}, 062001 (2016).

\bibitem{RevModPhys.72.545}
T. Brabec and F. Krausz, \textit{Rev. Mod. Phys.} \textbf{72}, 545 (2000).

\bibitem{DiMauro_et_al_JPhysB}
J. Xu, C. I. Blaga, P. Agostini, and L. F. DiMauro, \textit{J. Phys. B - At. Mol. Opt. Phys.} \textbf{49}, 112001 (2016).

\bibitem{PhysRevLett.71.1994}
P. B. Corkum, \textit{Phys. Rev. Lett.} \textbf{71}, 1994 (1993).

\bibitem{PhysRevA.49.2117}
M. Lewenstein, P. Balcou, M. Yu. Ivanov, A. L’Huillier, and P. B. Corkum, \textit{Phys. Rev. A} \textbf{49}, 2117 (1994).

\bibitem{jphysb47.204001}
S. V. Popruzhenko, \textit{J. Phys. B - At. Mol. Opt. Phys.} \textbf{47}, 204001 (2014).

\bibitem{Zuo1996313}
T. Zuo, A. D. Bandrauk, and P.B. Corkum, \textit{Chem. Phys. Lett.} \textbf{259}, 313 (1996).

\bibitem{JPhysB21.l31}
M. Ferray, A. L'Huillier, X. F. Li, L. A. Lompre, G. Mainfray, and C. Manus, \textit{J. Phys. B - At. Mol. Opt. Phys.} \textbf{21}, L31 (1988).

\bibitem{PhysRevA.48.R894}
B. Walker, E. Mevel, B. Yang, P. Breger, J. P. Chambaret, A. Antonetti, L. F. DiMauro, and P. Agostini, \textit{Phys. Rev. A} \textbf{48}, R894 (1993).

\bibitem{Meckel13062008}
M. Meckel, D. Comtois, D. Zeidler, A. Staudte, D. Pavi\ifmmode \check{c}\else\v{c}\fi{}i\ifmmode\acute{c}\else\'{c}\fi{}, H. C. Bandulet, H. P\'epin, J. C. Kieffer, R. D\"orner, D. M. Villeneuve, and P. B. Corkum. \textit{Science} \textbf{320} 1478 (2008).

\bibitem{Pullen2015}
M. G. Pullen, B. Wolter, A.-T. Le, M. Baudisch, M. Hemmer, A. Senftleben, C. D. Schr\"oter, J. Ullrich, R. Moshammer, C. D. Lin, and J. Biegert, \textit{Nat. Commun.} \textbf{6}, 7262 (2015).

\bibitem{nat.483.194}
C. I. Blaga, J. Xu, A. D. DiChiara, E. Sistrunk, K. Zhang, P. Agostini, T. A. Miller, L. F. DiMauro, and C. D. Lin, \textit{Nature} \textbf{483}, 194 (2012).

\bibitem{PhysRevA.85.053417}
M. Peters, T. T. Nguyen-Dang, E. Charron, A. Keller, and O. Atabek, \textit{Phys. Rev. A} \textbf{85}, 053417 (2012).

\bibitem{jphyschemA.102.4031}
D. Zhong and A. H. Zewail, \textit{J. Phys. Chem. A} \textbf{102}, 4031 (1998).

\bibitem{annurev-physchem-032210-103522}
W. Domcke and D. R. Yarkony, \textit{Ann. Rev. Phys. Chem.} \textbf{63}, 325 (2012).

\bibitem{nat.7.822}
C. Vozzi, M. Negro, F. Calegari, G. Sansone, M. Nisoli, S. De Silvestri, and S. Stagira, \textit{Nat. Phys.} \textbf{7}, 822 (2011).

\bibitem{nat.432.867}
J. Itatani, J. Levesque, D. Zeidler, Hiromichi Niikura, H. Pepin, J. C. Kieffer, P. B. Corkum, and D. M. Villeneuve, \textit{Nature} \textbf{432}, 867 (2004).

\bibitem{PhysRevLett.98.243001}
D. Pavi\ifmmode \check{c}\else\v{c}\fi{}i\ifmmode\acute{c}\else\'{c}\fi{}, K. F. Lee, D. M. Rayner, P. B. Corkum, and D. M. Villeneuve, \textit{Phys. Rev. Lett.} \textbf{98}, 243001 (2007).

\bibitem{PhysRevA.83.051403}
M. Peters, T. T. Nguyen-Dang, C. Cornaggia, S. Saugout, E. Charron, A. Keller, and O. Atabek, \textit{Phys. Rev. A} \textbf{83}, 051403 (2011).

\bibitem{0953-4075-27-7-002}
C. Cornaggia, M. Schmidt, and D. Normand, \textit{J. Phys. B - At. Mol. Opt. Phys} \textbf{27}, L123 (1994).

\bibitem{Elshakre201337}
M. Elshakre, \textit{Int. J. Mass. Spec.} \textbf{336}, 37 (2013).

\bibitem{doi:10.1021/ja0344819}
Y. Sato, H. Kono, S. Koseki, and Y. Fujimura, \textit{J. Am. Chem. Soc.} \textbf{125}, 8019 (2003).

\bibitem{QUA:QUA51}
S. Chelkowski and A. D. Bandrauk, \textit{Int. J. Quant. Chem.} \textbf{60}, 1685 (1996).

\bibitem{Lehtovaara2007148}
L. Lehtovaara, J. Toivanen, and J. Eloranta, \textit{J. Comput. Phys.} \textbf{221}, 148 (2007).

\bibitem{PhysRevA.52.1450}
A. Keller, \textit{Phys. Rev. A} \textbf{52}, 1450 (1995).

\bibitem{RevScIns86.4926994}
S. P. Weathersby, G. Brown, M. Centurion, T. F. Chase, R. Coffee, J. Corbett, J. P. Eichner, J. C. Frisch, A. R. Fry, M. G\"uhr, N. Hartmann, C. Hast, R. Hettel, R. K. Jobe, E. N. Jongewaard, J. R. Lewandowski, R. K. Li, A. M. Lindenberg, I. Makasyuk, J. E. May, D. McCormick, M. N. Nguyen, A. H. Reid, X. Shen, K. Sokolowski-Tinten, T. Vecchione, S. L. Vetter, J. Wu, J. Yang, H. A. D\"urr, and X. J. Wang, \textit{Rev. Sci. Instrum.}, \textbf{86}, 4926994 (2015).

\bibitem{RevModPhys.81.163}
F. Krausz and M. Ivanov, \textit{Rev. Mod. Phys.} \textbf{81}, 163 (2009).

\bibitem{Spanner:2004:12L02}
M. Spanner, O. Smirnova, P. B. Corkum, and M. Y. Ivanov, \textit{J. Phys. B - At. Mol. Opt. Phys} \textbf{37}, L243 (2004).

\bibitem{Meckel2014}
M. Meckel, A. Staudte, S. Patchkovskii, D. M. Villeneuve, P. B. Corkum, R. Dorner, and M. Spanner, \textit{Nat. Phys.} \textbf{10}, 594 (2014).

\bibitem{Chen:14}
L. Chen, C. Huang, X. Zhu, P. Lan, and P. Lu, \textit{Opt. Express} \textbf{22}, 20421 (2014).

\bibitem{Huismans07012011}
Y. Huismans, A. Rouz\'ee, A. Gijsbertsen, J. H. Jungmann, A. S. Smolkowska, P. S. W. M. Logman, F. L\'epine, C. Cauchy, S. Zamith, T. Marchenko, J. M. Bakker, G. Berden, B. Redlich, A. F. G. van der Meer, H. G. Muller, W. Vermin, K. J. Schafer, M. Spanner, M. Yu. Ivanov, O. Smirnova, D. Bauer, S. V. Popruzhenko, and M. J. J. Vrakking, \textit{Science} \textbf{331}, 61 (2011).

\bibitem{ATI_top_rev}
D. B. Milo\ifmmode \check{s}\else\v{s}\fi{}evi\'c, G. G. Paulus, D. Bauer, and W. Becker, \textit{J. Phys. B - At. Mol. Opt. Phys}, \textbf{39}, R203 (2006).

\bibitem{Frasca2195}
M. Frasca, \textit{Proc. R. Soc. London A} \textbf{463}, 2195 (2007).

\bibitem{Reiss_SFA}
R. H. Reiss, \textit{Progress in Ultrafast Intense Laser Science III}, Chapter Foundations of the Strong-Field Approximation, Pages 1--31, Springer, Berlin (2008).

\bibitem{PhysRevA.74.063404}
D. B. Milo\ifmmode \check{s}\else\v{s}\fi{}evi\'c, \textit{Phys. Rev. A} \textbf{74}, 063404 (2006).

\bibitem{NDSenGupta}
N. D. Sengupta, \textit{Bull. Calcutta Math. Soc.} \textbf{44}, 175 (1952).

\bibitem{Suarez_2015}
N. Su\'arez, A. Chac\'on, M. F. Ciappina, J. Biegert, and M. Lewenstein, \textit{Phys. Rev. A} \textbf{92}, 063421 (2015).

\bibitem{PhysRevA.78.033412}
M. Busulad\ifmmode\check{z}\else\v{z}\fi{}i\'c, A. Gazibegovi\'c-Busulad\ifmmode\check{z}\else\v{z}\fi{}i\'c, D. B. Milo\ifmmode\check{s}\else\v{s}\fi{}evi\'c, and W. Becker, \textit{Phys. Rev. A} \textbf{78}, 033412 (2008).

\bibitem{PhysRevA.54.742}
M. Yu. Ivanov, T. Brabec, and N. Burnett, \textit{Phys. Rev. A} \textbf{54}, 742 (1996).

\bibitem{Algorithm}
K. Levenberg, \textit{Quart. Appl. Math.} \textbf{2}, 164 (1944) ; D. Marquardt, \textit{J. Soc. Indust. Appl. Math.} \textbf{11}, 431 (1963).

%%%%%%%%%%%%%%%%%%%%%%
\end{thebibliography}
\end{document}